%% file: main.tex
\documentclass[%
reprint,
 amsmath,amssymb,
 aps,
prd,
showkeys,
]{revtex4-2}

\usepackage{sidecap}
\usepackage{graphicx}
\usepackage{dcolumn}
\usepackage{bm}

\newcommand{\argon}[0]{$^{37}$Ar }
\newcommand{\radon}[0]{$^{220}$Rn }

\usepackage{xcolor}

\begin{document}

\preprint{APS/123-QED}

\title{Investigating the XENON1T low-energy electronic recoil excess using NEST}

\author{M. Szydagis}\email{mszydagis@albany.edu}
\author{C. Levy}\email{clevy@albany.edu}
\author{G. M. Blockinger}
\author{A. Kamaha}
\author{N. Parveen}
\author{G.R.C. Rischbieter}

\affiliation{Department of Physics, University at Albany, State University of New York, Albany 12222-0100, New York, USA}

\date{\today}

\begin{abstract}

The search for dark matter, the missing mass of the Universe, is one of the most active fields of study within particle physics. The XENON1T experiment recently observed a 3.5$\sigma$ excess potentially consistent with dark matter, or with solar axions. Here, we will use the Noble Element Simulation Technique (NEST) software to simulate the XENON1T detector, reproducing the excess. We utilize different detector efficiency and energy reconstruction models, but they primarily impact sub-keV energies and cannot explain the XENON1T excess. However, using NEST, we can reproduce their excess in multiple, unique ways, most easily via the addition of 31~$\pm$~11 \argon decays. Furthermore, this results in new, modified background models, reducing the significance of the excess to $\le2.2\sigma$ at least using non-Profile Likelihood Ratio (PLR) methods. This is independent confirmation that the excess is a real effect, but potentially explicable by known physics. Many cross-checks of our \argon hypothesis are presented.

\end{abstract}


\maketitle

\section{\label{sec:intro}Introduction}
\vspace{-5pt}
There is overwhelming evidence, via astrophysical and cosmological observations~\cite{Rubin_2000,planck2018}, that the Universe is made of nonluminous matter interacting rarely with baryons. The search for the aptly-named ``dark matter'' has been an active field for decades. Experiments have been looking for different types, particularly weakly interacting massive particles (WIMPs) via direct nuclear recoils (NRs) and/or electronic recoils (ERs). While no experiment has made an unambiguous conclusive detection of dark matter or of axions~\cite{PecceiQuinn} that has not already been contested and/or explained, the newest results from the XENON1T experiment~\cite{Aprile:2020tmw} do exhibit an excess over their background for low-energy ER. While XENON1T was built to look predominantly for WIMPs, it is sensitive to the axion via ER, particularly solar axions, one potential explanation for the reported excess. For this work, we will not study potential solar axion detection, nor a neutrino magnetic moment or bosonic WIMPs. Instead, using the Noble Element Simulation Technique (NEST) software~\cite{NESTsoftware}, we focus on independently confirming a real excess, then seek alternate explanations.

Liquid xenon (LXe) detectors such as XENON1T need to be simulated with high precision, as in all rare event searches, before potentially new physics can be properly identified. While XENON1T has its own Monte Carlo (MC) framework~\cite{PhysRevD.99.112009}, whose advantage is in simulating features unique to the detector, the publicly available NEST simulation software is a toolkit that is widely used in the LXe community, and whose development team includes members of the LUX/LZ, XENON1T/nT, (n)EXO, and DUNE experiments. NEST has served numerous noble-element-based experiments during the nine years since its inception~\cite{NEST2011}, proving that it can accurately simulate and reproduce the results of various LXe (and liquid argon) detectors~\cite{LUX1,LZ,PANDAX,XENON}, by incorporating the immense amount of data available from calibrations and backgrounds (BGs). 
\vspace{-15pt}

\section{\label{sec:nest}Noble Element Simulation Technique}
\vspace{-10pt}
In a detector-agnostic way, NEST is capable of modeling average yield, i.e., numbers of quanta (photons or electrons) produced per unit energy, by various types of interactions: NR, ER, $\alpha$, $^{83m}$Kr, and heavy non-Xe ion recoils like $^{206}$Pb~\cite{NEST2015,CutterTalk}. It is also capable of simulating detector specifics like energy resolution, both standard deviation of monoenergetic peaks and the widths of the log(S2) and log(S2/S1) ``bands'' (where S1 and S2 refer to the primary and secondary scintillation signals in noble elements). NEST can thereby simulate the leakage of ER events into the NR region and quantify the background discrimination in WIMP searches. In its simulating both the mean yields and resolution, NEST is able to model efficiencies, and so thresholds. We heavily take advantage of this capability in this work. Lastly, NEST can reproduce S1 and S2 pulse shape characteristics, but they are unneeded here except for the S1 coincidence window.

\begin{figure}
\begin{center}
\includegraphics[width=0.44\textwidth,clip]{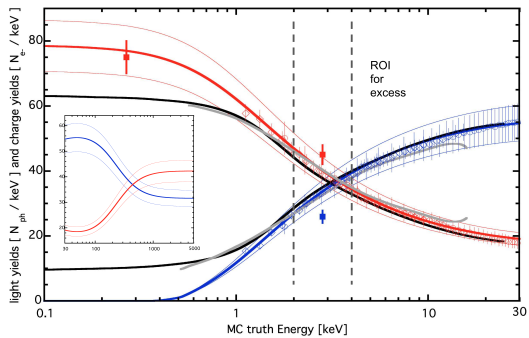}
\vspace{-10pt}
\caption{NESTv2.1 $L_{y}$ (blue), $Q_{y}$ (red) for betas at 81~V/cm. Bands represent $\pm$10\%, a typical estimate of the systematic uncertainty in NEST, driven primarily by uncertainties in S1 and S2 gains in the data ($g_{1}$ and $g_{2}$)~\cite{LUXg1g2}. XENON100's $^{3}$H-based-model is in gray, with XENON1T's~\cite{PhysRevD.99.112009} in black using \radon at the closest E-fields with which we can compare, 90 and 82~V/cm, respectively~\cite{XENON3H}. The circles and diamonds are 80~V/cm $^{14}$C and $^{3}$H LUX data sets, respectively~\cite{Akerib:2019jtm}, while squares are \argon data from PIXeY~\cite{pixey} at $\sim$100~V/cm. One reason for the slight discrepancy is the $L_{y}$ increasing ($Q_{y}$ anti-correlated) with lower drift field. $Inset$: yields out to 5~MeV.}
\vspace{-24pt}
\label{yields}
\end{center}
\end{figure}

We reanalyze \cite{Aprile:2020tmw} here, utilizing NEST to try and explain excess events as being, e.g., from an unexpected BG. NEST average yield and width parameters did not need to be varied to fit to XENON1T data, as they are detector-independent. Only the detector-specific values were changed to match XENON1T. This is made clear in Fig.~\ref{yields}. At sub-keV energies, light yield goes to 0, as, in opposite fashion, charge asymptotes to its maximum possible value, with NEST uncertainty spanning the possibilities ranging from taking the inverse of the ``traditional'' $W$ value of 13.7~$\pm$~0.2~eV~\cite{NEST2015} (73 quanta/keV) to the reciprocal of the recent measurement from EXO, 11.5~$\pm$~0.5~eV (i.e., 87 quanta/keV)~\cite{EXO200WValue}. However, in the region of greatest interest for our analysis, indicated by vertical dashes in Fig.~\ref{yields}, the default NEST yields MC simulation for electrons is in outstanding agreement with all the existing relevant data sets and models. Disagreement at energies orders of magnitude away from this region of interest (ROI) is less relevant, but also still small (Fig.~\ref{yields} inset).

It is therefore no surprise we find NEST able to ``postdict'' the XENON1T results at 81~V/cm without any free parameters. This occurred despite the fact that there is less calibration data at this low drift field (compared to past experiments operated at $O$(100--1000)~V/cm) upon which to base NEST's low-field yields model for ER: $L_{y}$ (photons/keV) and $Q_{y}$ ($e^{-}$/keV). So, we were able to use Fig.~\ref{yields}'s central red and blue lines, without floating yields.

It is also worth noting that, despite there being a recent new stable release of NEST, the beta yield model has not been officially updated in over two years. Recent LUX work with a $^{14}$C beta source~\cite{Akerib:2019jtm} is not the default but instead a NEST option, to avoid potential overfitting to LUX at the expense of earlier global data. The default NEST yields applied in this work were fit to LUX tritium data but not to the LUX $^{14}$C data. NEST was never used for a \radon calibration before now, being driven primarily by tritium, yet it works successfully, as will be seen next.
\vspace{-22pt}

\section{Methods}
\vspace{-7pt}
The primary method employed here is simple: we first reproduce XENON1T's calibration data, striving to understand their energy resolution, detector efficiency, and background model. We simulated data taken under the conditions of their experiment in NEST, and then compared that output to official XENON1T results.

For NEST to accurately and precisely simulate a detector, the first key input involves a proper detector parameter file. For complete transparency, Table~\ref{detparam} defines all parameters used as input to NEST that can be found publicly, except for the precise dimensions of the fiducial volume, which were set in NEST to best reproduce the fiducial mass of 1042$\pm$12~kg. The most important values NEST must have are $g_{1}$, $g_{2}$, and the drift electric field.

\vspace{-8pt}
\begin{table}[h!]
\centering
\begin{tabular}{|m{5.5cm}|m{2.7cm}| } 
 \hline
 \multicolumn{2}{|c|}{\textbf{Primary scintillation (S1) parameters} } \\
  \hline
  $g_1$ [phd/photon] & 0.13 \cite{AxionTalk} \\
  Single photoelectron resolution & 0.4 \cite{Behrens}\\
  Single photoelectron threshold [phe] & 0 (*eff used) \\
  Single photoelectron efficiency* & 0.93 \cite{Aprile:2018dbl}\\
  Baseline noise & 0 (assumed small)\\
  Double phe emission probability & 0.2 $\pm$ 0.05 \cite{Aprile:2017aty,LOPEZPAREDES201856}\\
  \hline
  \multicolumn{2}{|c|}{ \textbf{Ionization or secondary scintillation (S2)} } \\
  \hline
  $g_1^{gas}$ [phd/photon] & 0.1  \cite{AxionTalk, Aprile:2018dbl}\\
  Single $e^{-}$ (SE) size Fano-like factor & 1.0\\
  S2 threshold [phe] top + bottom & 500 (uncorr) \cite{Aprile:2020tmw}\\
  Gas extraction region field [kV/cm]  & 10.8 (est.) \cite{Aprile:2018dbl}\\
  Electron lifetime [$\mu$s] & 650  \cite{Aprile:2018dbl}\\
  \hline
   \multicolumn{2}{|c|}{\textbf{Thermodynamics properties}} \\
  \hline
  Temperature [K] & 177.15  \cite{Aprile:2018dbl}\\
  Gas pressure [bar] & 1.94 (abs) \cite{Aprile:2018dbl}\\
  \hline
  \multicolumn{2}{|c|}{ \textbf{Geometric and analysis parameters} } \\
  \hline
  Minimum drift time [$\mu$s] & 70 \cite{Aprile:2018dbl}\\
  Maximum drift time [$\mu$s] & 740 \cite{Aprile:2018dbl}\\
  Fiducial radius [mm]& 370 \cite{Aprile:2020tmw, Aprile:2018dbl}\\
  Detector radius [mm]& 960  \cite{Aprile:2017iyp}\\
  LXe-GXe border [mm]& 1031.5  \cite{Aprile:2017iyp}\\
  Anode level [mm]& 1034  \cite{Aprile:2017aty}\\
  Gate level [mm]& 1029  \cite{Aprile:2017aty}\\
  Cathode level [mm]& 60 \cite{Aprile:2017aty}\\
 \hline
\end{tabular}
\caption{Summary of XENON1T detector parameter values implemented for NEST in this work. Please note the $g_{1}$ does not match a published value, as standard phe units include the 2-phe effect (whereby one VUV photon can make 2 phe within a PMT~\cite{2PE}). We therefore quote a different $g_{1}$, in our style of detector modeling, using the unit of ``phd'' (detected photons) developed by LUX~\cite{LUXPhD}, with the 2-phe effect separately simulated, probabilistically (not a constant offset) as done also by XENON1T. Lastly, in NEST $z=0$ (the vertical axis) is at bottom, requiring a translation from XENON1T's definition, of $z=0$ at the top (gate grid wires).}
\label{detparam}
\vspace{-8pt}
\end{table}

We further assumed a threefold coincidence requirement, across 212 active PMTs (Photomultiplier Tubes), applying a 50.0~ns coincidence window~\cite{Aprile:2017aty}. Based on all of these inputs, NEST will output a $g_2$ (an emergent property based on gas light collection, extraction, and other separate effects modeled from first principles~\cite{LUXGreg}) of 9.85 phd/$e^-$ (or, 11.57 phe/$e^-$). This can be separated into an electron extraction efficiency of 95\%, derived from PIXeY/LLNL~\cite{pixey2,llnl}, and an underlying SE = 10.37~phd/$e^-$ = 12.18~phe/$e^-$. In using Poissonian statistics, we modeled a SE (1$\sigma$) width of 3.2 phe/$e^-$. The pressure and temperature reported lead to a simulated density of 2.86~g/mL and ($e^-$) drift speed of 1.26~mm/$\mu$s, a velocity which does appear to make the physical coordinates of their reported detector geometry match with the min and max drift times of the fiducial volume. The density also leads to an expected $W$ = 13.5~eV according to NEST (which models the work function for creation of quanta as being dependent on density, including across phases) which conveniently splits the difference between the Dahl and neriX values of 13.4--13.7~eV \cite{Dahl:2009nta,Goetzke:2016lfg}. This is a very small effect, however, and an overall scaling of $O$(1\%). It is therefore a negligible systematic.

\subsection{Energy resolution}
\begin{figure}
\begin{center}
\includegraphics[width=0.54\textwidth,clip]{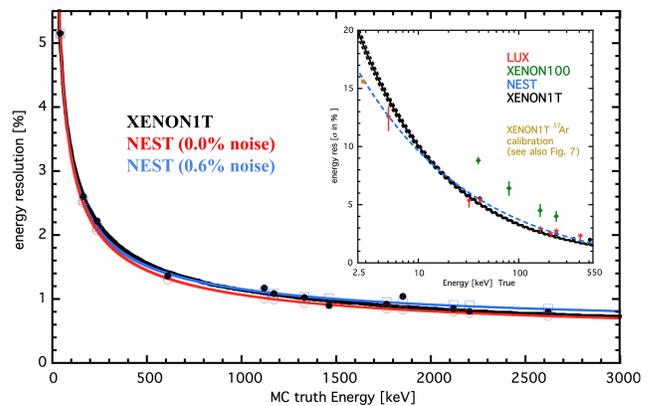}
\caption{Energy resolution~vs.~energy, comparing black dots, real data from XENON1T~\cite{Aprile:2020yad-ERes}, to NEST with 0\% noise (hollow red circles) and with 0.6\% noise (cyan squares). Lines are analytic fits (power laws plus constants, with powers consistent with the theoretical 0.5).  Black line is XENON1T model. $Inset$: the resolution at lower energy (down to 2.5~keV) with XENON1T's empirical function extrapolated from higher energies in black~\cite{Aprile:2019dec}. NEST with 0.006 noise is the cyan dash, extending the same simulations from the primary figure. Once $g_{1}$, $g_{2}$, and E-field are established, they drive the resolution in this energy range, from first principles. Data sets from other experiments are displayed as points, in other colors, but are not expected to match as resolution is unique per experiment. The yellow square in the inset will be addressed later.}
\vspace{-20pt}
\label{Eres}
\end{center}
\end{figure}

\vspace{-10pt}
We confirmed the veracity of detector parameters and the fluctuation model, covering both correlated and anti-correlated noise, by verifying NEST's predicted resolution for XENON1T as a function of energy~\cite{Aprile:2020yad-ERes} in Fig.~\ref{Eres}. This reveals that the ``linear'' noise, set by default (unrealistically) to 0.0 in NEST is closer to 0.6\%. Even without the addition of noise, the energy resolution predicted by NEST (without free parameters) is in good agreement with the XENON1T data; this implies that XENON1T achieved extremely low levels of noise and other effects, not captured within NEST by default. The difference is $<$~1\% (relative) comparing to results with/without noise. It is modeled as additional, uncorrelated Gaussian smearing and applied separately to the S1 and S2 pulses; it is directly proportional to each of the pulse areas in phe.

This accounts for imperfect position-dependent light collection, field uniformity, liquid leveling, plus similar known and unknown effects. Typical linear noise values, even given high-statistics $^{83m}$Kr and/or $^{131m}$Xe calibrations for efficiency and field mapping, are $\sim$1\%--4\%, with near-identical values for S1 and S2 (given the same DAQ being used for all pulse types) whenever NEST is used to match the past world data from different experiments \cite{Dahl:2009nta,LUXAgain}. We do set the noise to 0.6\% here, as it appears to create a better match to XENON1T, particularly at lower energies, as shown in Fig.~\ref{Eres}. Nevertheless, we have effectively performed an unbiased side-band calibration of the noise level here, as the lowest data point within Fig.~\ref{Eres} is at 41.5~keV, but the solar axion signal model does not extend beyond 30~keV~\cite{Aprile:2020tmw}.

\begin{figure}
\begin{center}
\includegraphics[width=1.04\textwidth,clip,angle=270]{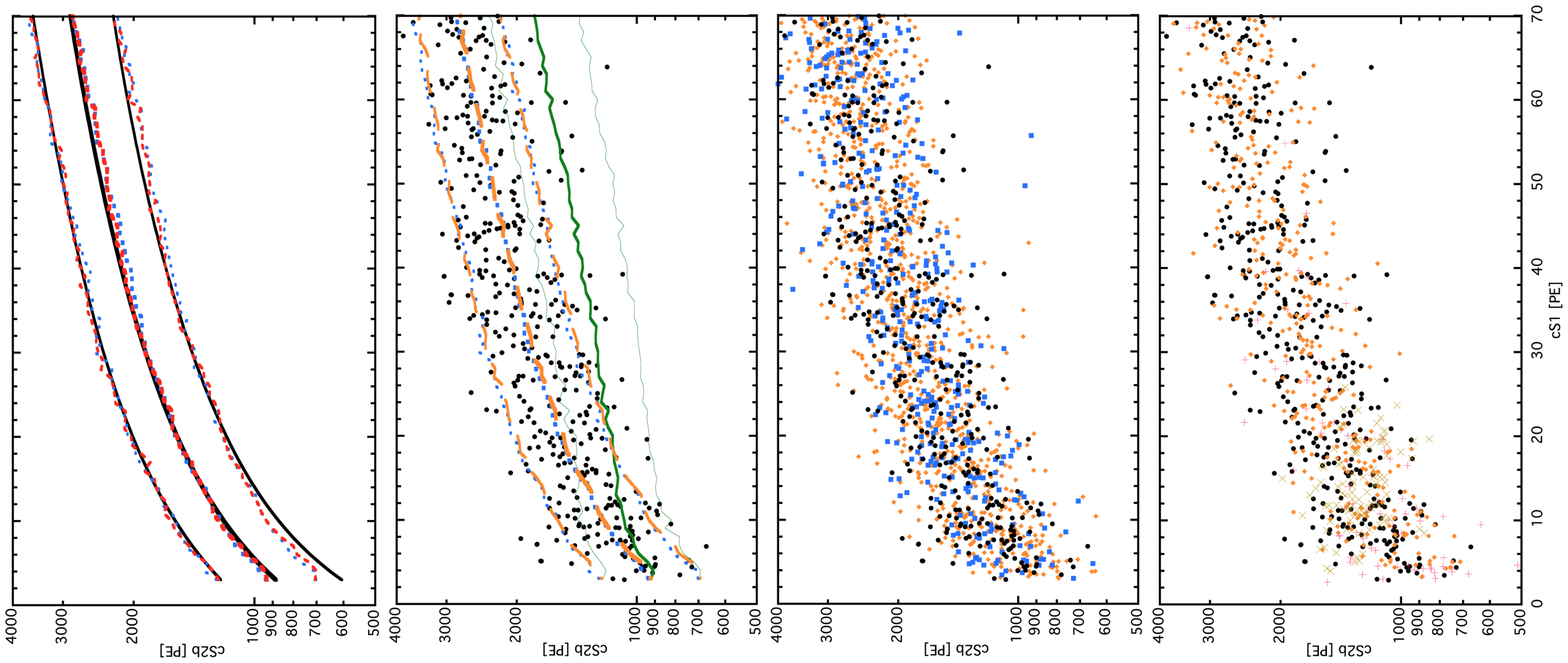}
\vspace{-15pt}
\caption{Top: NEST in dashed red reproducing high-statistic \radon calibration superimposed in solid black~\cite{Aprile2019:Rn220Fig16}, with skew-Gaussian fits, compared to flat BG (dashed cyan). Points for black not provided but can be seen in~\cite{Aprile2019:Rn220Fig16}. Contours indicate 10--50--90\%. Second: as both the BG and \radon are $\sim$flat that sim is repeated again in cyan, but now our custom generator appears (orange) and a generic flat gamma band (solid green). Science data from \cite{Aprile:2020tmw} as black dots. Third: NEST scatter plot overlaid on the XENON1T BG: a flat model (cyan squares) and custom generator (orange diamonds). XENON1T search data as black points again. Bottom: repeating orange from last plot, but fewer events, and adding yellow X's (\argon) and pink pluses (exponential BG), potential excess explanations.}
\vspace{-30pt}
\label{AllS2vS1Bands}
\end{center}
\end{figure}

\vspace{-10pt}
\subsection{NEST reproduction of the \radon calibration}
\vspace{-5pt}
To further confirm NEST simulates XENON1T well, we validate it against \radon data. We simulate 10$^{7}$ $^{212}$Pb beta decays that dominate~\cite{Lang:2016zde} as well as a flat (i.e., uniform in energy) spectrum, as the $^{212}$Pb is close to flat. Figure~\ref{AllS2vS1Bands} top compares with both. This demonstrates we reproduce \radon while Fig.~\ref{AllS2vS1Bands} second from top potentially explains outliers in \cite{Aprile:2020tmw} as due to gamma/x-rays, as they have different yields compared with betas at this energy scale~\cite{NEST2013}. Our hypothesis can also explain why this type of event is seen in BG data, but not $^{3}$H/$^{14}$C calibrations in XENON100/LUX. However, these may be gamma-X/MSSI (multiple-scatter single-ionization) BGs, possibly more insidious in this higher S1 range up to 70~phe, as opposed to 20--50 phe in earlier experiments~\cite{GregGammaX}. Detector geometry plays a strong role in gamma-X.

The flat ER BG spectrum shows that even in this crude way we still reproduce XENON1T well. To be quantitative, we compare not the flat MC but Rn MC with data. $(NEST-data)/data$ (NEST is red in Fig.~\ref{AllS2vS1Bands} top) has a median offset from the data (black line~\cite{Aprile2019:Rn220Fig16}) of --1.0\% for band mean, with (nonsystematic) max/min deviation of $\pm$5\%. For band width, the median offset is +1.3\% with max/min $\pm$12\%. This is quite comparable to what can be achieved with NEST with direct access to data~\cite{LUXGreg}.

\vspace{-5pt}
\subsection{XENON1T ER background NEST generator}
\vspace{-5pt}
Of equal importance to reproduction of the \radon calibration is BG generation, for obtaining simulated points: orange, in lower half of Fig.~\ref{AllS2vS1Bands}, contrasted with flat in cyan and data in black. A custom generator was created to follow the XENON1T ER BG model, corrected for detection efficiency, below 30 keV, allowing for a significant buffer beyond the excess ROI. By not including detector efficiency initially, we ensure the generator inputs the ``true energies'' into NEST, as an unadulterated, uncorrected energy spectrum, independent of detector effects.

\begin{figure}
\begin{center}
\includegraphics[width=0.475\textwidth,clip]{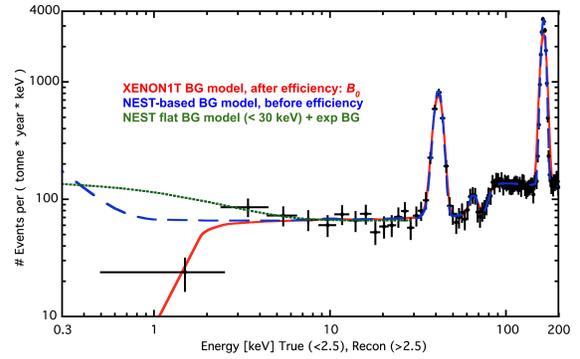}
\vspace{-5pt}
\caption{Smooth depiction of BGs assumed by NEST in blue (long dash), compared to B0 in red (solid), and observed data in black (points with errors)~\cite{Aprile:2020tmw}. The exponential is in green (dotted) that best fits the excess (no \argon necessary). Given uncertainties, including systematics in B0 and in efficiency, not depicted, it might be possible to reconcile the blue with the green. Figure~\ref{PinkEff} implies that is not necessarily the case, but a combination may be possible where the number of necessary \argon events to explain the data is dropped, reducing the steepness required in the green to explain the excess. (Note green becomes pink and blue becomes orange in other plots.)}
\vspace{-25pt}
\label{ExpRise}
\end{center}
\end{figure}

NEST's PDF is the sum of all XENON1T BGs in Table~1 of \cite{Aprile:2020tmw}, which includes the mean rate for each isotope fit by XENON1T, and the energy spectrum shapes assumed. Our check of the excess remains independent, as the use of NEST instead of XENON1T MC leads to some variations in energy resolutions, as seen in Fig.~\ref{Eres} and next in Fig.~\ref{ExpRise} where the 163.9~keV peak ($^{131m}$Xe) differs slightly. The sum of all BGs is indistinguishable statistically (Kolmogorov-Smirnov (KS) test, quoted below) from flat due to fluctuations, prior to addition of $^{37}$Ar. See Fig.~\ref{AllS2vS1Bands} for scatter, Fig.~\ref{excess} for energy binning.

In Fig.~\ref{ExpRise}, we explicitly show what the XENON1T BG looks like before efficiency. It is quite flat for $\sim$4--30 keV, but has a slight positive slope, after all radioisotopes are combined that contribute, which we do not neglect. Computing this was a necessary initial step. This is not in \cite{Aprile:2020tmw} but was derived by combining all isotopic contributions and dividing by the efficiency. The resultant shape better motivates qualitatively our investigation later into a BG that rises as energy goes to zero.

To show our generator functions, we simulate BG with it, and compare the outputs to data along with our first simplified flat model once again. 1-D unbinned KS tests in both S1s and S2s, running the generator repeatedly with different seeds on different systems and with different events counts (both greater than and equal to the 409 real points) produced p-values of 0.1--0.3 with both models, without a consistent improvement when applying noise, as small pulse areas are less affected by it.

These p-values are not indicative of any significant degree of statistical inconsistency. The reason they are not uniformly distributed up to 1.0 is likely the divergence at the lower half of the band for the lowest S1s, most easily observed in Fig.~\ref{AllS2vS1Bands} (top). This issue is not challenging to understand, but difficult to model without access to all information on XENON1T. It is likely due to a combination of wall and accidental-coincidence BGs that are not included in NEST. This feature can be observed in both \cite{Aprile:2020tmw} and Fig.~16 of \cite{Aprile2019:Rn220Fig16}. Looking at the low-S1 upper half of the band, it is clear this is a problem with the symmetry of the band, but not with band width overall. Alphas on the wall or from \radon itself degraded in energy, as well as heavy recoiling atoms/ions such as Pb from the Rn chain and naked betas, may experience charge loss, lowering the S2. As this is a problem only at lower S2s, and appears below the average S1 for $^{37}$Ar, our results stand in spite of this, but this is a problem even in the fiducial volume far from the walls: degraded position resolution at walls can cause a low S2 to be reconstructed inside of the fiducial volume. For higher S2s, apparent leakage of events above the upper Gaussian contour can be ascribed to multiple-scatter ER identified as single-scatter and/or Xe-intrinsic upward skew, as detailed later for peaks.

\vspace{-15pt}
\subsection{Energy reconstruction and efficiency}
\vspace{-5pt}

The excess was measured in binned energy space not S2 versus S1 scatter, so that defined the next investigation. XENON1T reports reconstructed energies, but the non-linear deconvolution into true energy was estimated via MC~\cite{PhysRevD.99.112009,Aprile2019:Rn220Fig16} for their PLR. There may be differences from NEST, but primarily at sub-keV, thus irrelevant.

The efficiency was verified many ways, but again NEST agrees with what was reported except sub-keV, not relevant here, and also within large errors (in the Appendix).

\vspace{-15pt}
\section{Results}
\begin{figure}
\begin{center}
\includegraphics[width=0.8\textwidth,angle=270]{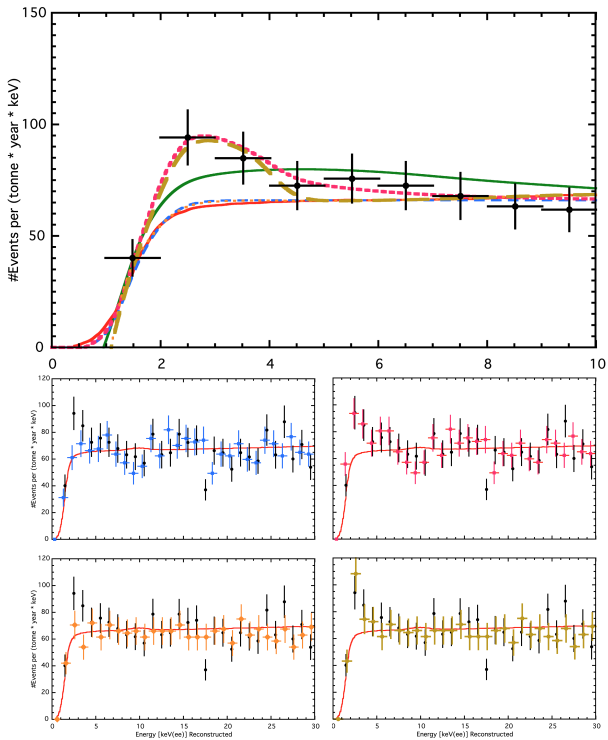}
\vspace{-130pt}
\caption{A summary of every model studied with NEST: data and background B0 model from \cite{Aprile:2020tmw} are black dots and solid red line, respectively. Top: our flat ER BG (cyan), the same flat ER BG with a low-energy exponential added (pink), the NEST custom generator for mimicking B0 (orange), the same custom generator with \argon (yellow), then with tritium added (thick solid green line). Bottom: the discrete NEST outputs in the same colors as at top, but after realistic full, detector MC. For clarity, every point has been offset from its actual value by $O$(0.1)~keV and $^3$H is omitted. The flat BG is 66 per tonne-year-keV. Random seed used is identical per each row.}
\vspace{-20pt}
\label{excess}
\end{center}
\end{figure}

\vspace{-10pt}
The NEST-simulated energy histograms are depicted in Fig.~\ref{excess}. The top only shows the region of interest below 10~keV but we explored up to 30~keV as shown at bottom. Black circles are always real data points as reported by \cite{Aprile:2020tmw}. We first modeled XENON1T's ER background using NEST, assuming a flat background (cyan squares), then using our custom generator (orange diamonds again).

The difference between ``B0,'' the XENON1T BG model after efficiency application in red, and the other curves near 1~keV in Fig.~\ref{excess} is due to NEST's lower intrinsic efficiency, as predicted based on $g_{1}$, $g_{2}$, and field, but this (insignificant) disagreement is far from the ROI. However, \argon does fall well within the ROI and, based also on LUX experiences~\cite{attilathesis,LUXAnnualMod}, is our primary attempt to explain the excess. We at first added 50 \argon events over the full 0.65 tonne-year exposure, estimated from the raw size of the excess, later refined to 31~$\pm$~11 counts as best fit. \argon exhibits two low-$E$ peaks: 0.27 and 2.82~keV. While the latter is the one of interest here, as it may lie near the location of the excess in XENON1T's main analysis, the lower-$E$ peak may permit us to distinguish between \argon and other potential BGs. Our MC simulation corresponds to $48~^{+17}_{-18}$~\argon decays per tonne-year of exposure. Fig.~\ref{AllS2vS1Bands} bottom shows them in S2~vs.~S1.

\begin{figure}[b]
\begin{center}
\includegraphics[width=0.45\textwidth,clip]{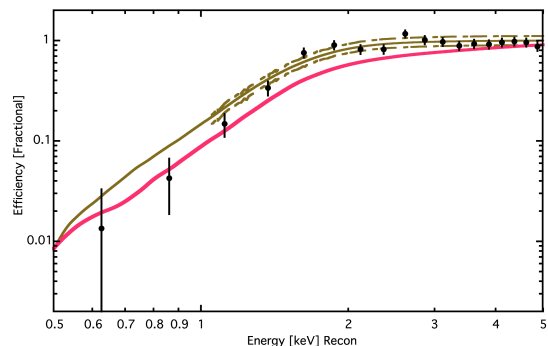}
\vspace{-10pt}
\caption{XENON1T efficiencies from Figs.~2 and 6 of \cite{Aprile:2020tmw} copied here in mustard (with error) and black (solid circles), where \radon was used to calculate the latter. Pink shows what efficiency would be to justify an exponentially falling BG.}
\vspace{-20pt}
\label{PinkEff}
\end{center}
\end{figure}

We also model an exponential background added to a flat ER background (pink in Fig.~\ref{excess}). However, it is not motivated by a specific new BG physically. It is purely mathematical, but shows that adding either a spectrum, or a monoenergetic peak, can reproduce the excess. As the flat~+~exponential model fits so well, we try to motivate the excess by an underestimation in the BG model, via an overestimation of efficiency. However, we find the efficiency would have to be over 2$\sigma$ off for several data points in a row, in the ROI, to justify such a drastically different BG, as shown in Fig. \ref{PinkEff}. This is less compelling.

Lastly, we model tritium ($^3$H), but also find it to be less compelling. It is not only a worse fit than \argon and the exponential (if you account for shape using $\chi^2$, and do not just look at Poisson statistics), it is lower than the other hypotheses in the 2.5~keV bin, farther from the data. It also raises the counts in the lowest energy bin due to this being a continuous source, unlike \argon which is monoenergetic. The exponential hypothesis suffers less from this raising of counts for the 1.5~keV bin considerably above the data, as, counterintuitively, exponentially more counts at low energies implies more counts at true energies which are unable to fluctuate up effectively, in \textit{reconstructed} energy space. We fully recognize these statements could be strengthened with a PLR, but without access to all data in all dimensions including position this is unrealistic for non-XENON1T members.

Table~\ref{chi2} has the $\chi^2$'s and the $\sigma$ discrepancies between our models and the data points (black dots from Fig.~\ref{excess}). For completeness, and to reproduce the XENON1T numbers, we considered the 1--7~keV range. However, due to the size of the error bars, we find that the fits, and thus $\chi^2$'s, are overconstrained over this range. Therefore, we choose to fit to a larger energy range (1--30~keV, as per Fig.~4 of~\cite{Aprile:2020tmw}). This shows that our best fit to the data is using an exponential BG, followed by \argon then tritium.

\begin{table}
    \centering
    \begin{tabular}{|c|c|c|c|c|c|}
    \hline
    \multicolumn{6}{|c|}{ \textbf{1--30~keV ($\sigma_{\chi}$) and 1--7~keV ($\sigma_{p}$)} } \\
     \hline 
     Hypothesis (color) & $\chi^2$ & d.o.f & $\chi^2$/d.o.f & $\sigma_{\chi}$ & $\sigma_{p}$ \\
     \hline
    Flat BG (cyan) &41&29-1& 1.46 & 1.92 & 2.65 \\
    B$_{0}$ (red) &48&29-4&1.92&2.91 & 3.35 \\
    PDF (orange) &47&29-4&1.88&2.80 & 2.70 \\
    PDF + \argon (yellow)&38&29-5&1.60&2.16 & 0.41 \\
    Flat + exponential (pink) &33&29-3&1.26&1.38 & -0.54 \\
    PDF + $^3$H (green) &45&29-5&1.88&2.80 & -0.28 \\
    \hline
    \end{tabular}
    \caption{The goodness of fit quantifying the level of agreement with data, for two broad energy ranges (encompassing the 2--4~keV range where the excess seems largest). The number of free parameters assumed for B0 and the PDF, our custom B0-like generator, is four, representing the four largest low-energy BGs ($^{214}$Pb, $^{85}$Kr, solar $\nu$'s, materials) and three for the exponential (amplitude, shape, offset). With $^{3}$H and $^{37}$Ar, one additional parameter was varied for the PDF, the number of decays. Flat had only one free parameter. Using na\"{\i}ve counting, in a tighter energy window, all excess hypotheses do well, as denoted by $\sigma_{p}$ (p = Poisson). $\sigma_{\chi}$ is derived from the $\chi^2$/d.o.f (degrees of freedom).}
    \label{chi2}
    \vspace{-15pt}
\end{table}

\begin{figure}
\begin{center}
\includegraphics[width=0.335\textwidth,clip]{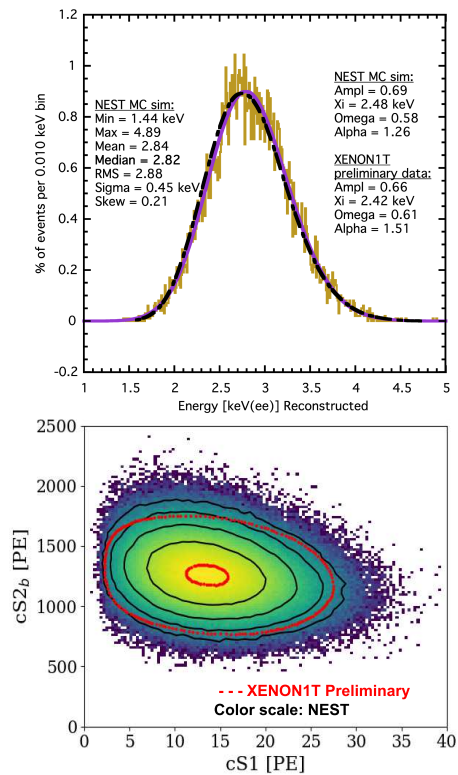}
\vspace{-8pt}
\caption{Top: \argon peak. NEST utilizing XENON1T detector parameters in gold, best fit (skew-normal) in magenta. Black dash is a preliminary XENON1T calibration~\cite{AxionTalk} showing again remarkable agreement with (default) NEST. Numbers at left are raw histogram statistics; at right best-fit parameters, defined on next page, for both NEST and data, with errors $\sim$0.1 in each due to high statistics. Bottom: S2 versus S1 for $10^{6}$ \argon events. The color scale and black contours are both NEST's; in red are inner/outermost (arbitrary) contours of slide~68 of \cite{AxionTalk}. As raw data were marked as preliminary, not provided by XENON1T, only a qualitative comparison can be performed.}
\vspace{-19pt}
\label{skewedgaus}
\end{center}
\end{figure}

\argon does not span 1.5--3.5~keV bins equally, when at 2.8~keV it should be $\sim$symmetric about 2.5~keV. This is due to positive skew (Fig.~\ref{skewedgaus}). At near-threshold energies, event triggering occurs on high-S1 tails. Moreover, skew in NEST enters at the level of recombination probability for S2 electrons, derived from LUX calibrations~\cite{VetriLUX}. It appears not only in ER bands but monoenergetic peaks. Figure~\ref{skewedgaus} shows the \argon 2.82~keV peak. A fit of the NEST histogram to a skew-normal distribution has skewness $\alpha$ (described later) of 1.3~$\pm$~0.2, compared with 1.5~$\pm$~0.2 for preliminary XENON1T calibration data. The skewness effect, already observed for \argon\cite{pixey,boulton}, is again not specific to it~\cite{VetriLUX}; the effect will be more prominent for monoenergetic peaks than for a broad spectrum of different energies like tritium, due to smearing.

In Fig.~\ref{skewedgaus} bottom, we use NEST to further study actual $^{37}$Ar, which was a XENON1T calibration, not just potential BG or excess hypothesis, affording us an opportunity of a deeper independent study. For this plot, we separate combined energy into the S1 and S2 areas. The non-Gaussian, triangular shape qualitatively agrees with data. This should make the probability of a NEST mismodeling of \argon in XENON1T impacting our result \textit{de minimis}. To allow additional, quantitative comparison, in combined-energy space, we quantify our work in Fig.~\ref{skewedgaus} top.

A similar asymmetry was in fact already reported by XENON1T: after discovering low outliers, lying below their ER band (Sec.~III~B: possibly $\gamma$'s and/or $\gamma$-X), not just high outliers above the band (as expected based upon the skew observed in their calibration bands), they added a BG ``mismodeling'' parameter into their WIMP search to compensate for any lower (i.e., subband) outliers~\cite{Aprile:2017iyp,ModelSafeGuard}. They did ultimately determine though that fewer WIMP-signal-like (NR-like) ER tail events in science data compared to calibration were a better fit~\cite{Aprile:2018dbl,PhysRevD.99.112009} and also provided an explanation for remaining outliers as being driven primarily by surface BGs, which experience charge loss, lowering their S2, similar to what was found on LUX~\cite{LUXComplete}, and mentioned earlier. We presented here a novel explanation that can perhaps account for a fraction of the outliers in Fig.~\ref{AllS2vS1Bands} (XENON1T's Fig.~5). Further evidence in favor of gammas is in the Appendix. They are not likely to explain all outliers as the NR band would be too contaminated for a WIMP search then.

\begin{figure}
\begin{center}
\includegraphics[width=0.48\textwidth,clip]{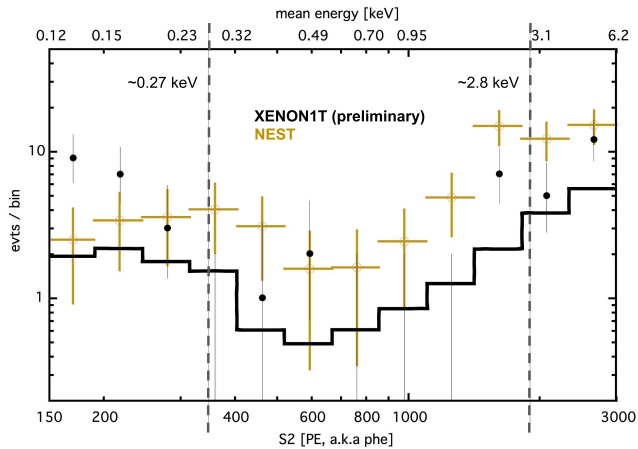}
\vspace{-15pt}
\caption{S2-only data from NEST (gold) simulated by adding the same amount of \argon as in the primary analysis to the preliminary XENON1T S2-only BG model~\cite{AxionTalk} (black steps). Excess over BG at $\sim$2000~phe (or photoelectrons (PE)) is consistent with 2.82~keV in NEST, consistent with the preliminary XENON1T data points~\cite{AxionTalk} (black dots). Errors on y are Poisson; on x, bin width. Lastly, while a na\"{\i}ve scaling (0.27/2.8) * 1900 = 180 would reproduce the first bin excess, the energy dependence of $Q_{y}$ does not justify that.}
\vspace{-25pt}
\label{S2only}
\end{center}
\end{figure}

Another important check upon the validity of the \argon hypothesis comes from looking at the S2-only analysis. Note that this will be in units of the total S2 signal, as opposed to bottom-PMT-array, and it is uncorrected, as the lack of S1 makes 3D position correction impossible. If the excess is due to $^{37}$Ar, then we expect additional excess at low S2s due to the 0.27 keV peak from the $^{37}$Ar, along with more events at high S2s due to the 2.82~keV peak. Our NEST simulation is compared to the XENON1T S2-only cross-check~\cite{AxionTalk} and it is shown in Fig.~\ref{S2only}. Within the statistics of the existing data provided by XENON1T, the S2-only analysis can neither rule out, nor rule in, the \argon hypothesis. It is not, however, inconsistent with it and can thus be the means to explain the excess event counts with respect to the S2-only BG model in most bins, even if they are not individually statistically significant.

The comparison at the lowest energy bins is less compelling, with the excess over BG occurring at lower S2 than simulated with NEST at 0.27~keV with the proper branching ratio. However, 
Fig.~\ref{yields} hints 
this could be explained 
within NEST's large uncertainties on $Q_{y}$ for this extreme low-energy regime. Furthermore, as this is uncorrected S2, we would need a full XY map and $e^{-}$--lifetime (vs.~time) to simulate XENON1T more precisely. Lastly, few-$e^-$ BGs from multiple sources, e.g., grid wire emissions~\cite{XENON}, may be coming into play for the first bin. Because of these enormous systematics, we do not pursue the S2-only avenue further, not considering, e.g., tritium.

A drawback to the \argon hypothesis is the best fit to a peak for a bosonic dark matter search being 2.3~keV: in XENON1T's Fig.~11 (\cite{Aprile:2020tmw} v2) 2.8~keV is strongly disfavored. We now reconcile our hypothesis with this analysis. XENON1T states more than once that the functional form used in \cite{Aprile:2020tmw} was Gaussian, so their peak search does not account for the inherent asymmetry due to skew at keV-scale energies (Fig.~\ref{skewedgaus} again) demonstrated by their own calibration, which they do not include in their analysis. The formulas for a skew-normal fit are as follows:

\vspace{-4pt}
\begin{flushleft}
(1) $y = A e^{\frac{-(x-\xi)^{2}}{2\omega^{2}}} [ 1 + erf ( \alpha \frac{x-\xi}{\omega \sqrt{2}} ) ]$ \\
(2) $\mu = \xi + \omega \delta \sqrt{\frac{2}{\pi}}$ \\
(3) $\delta = \alpha~/~\sqrt {~1 + \alpha^{2}~}$ \\
\end{flushleft}
\vspace{-1pt}

\begin{figure}
\begin{center}
\includegraphics[width=0.53\textwidth,clip]{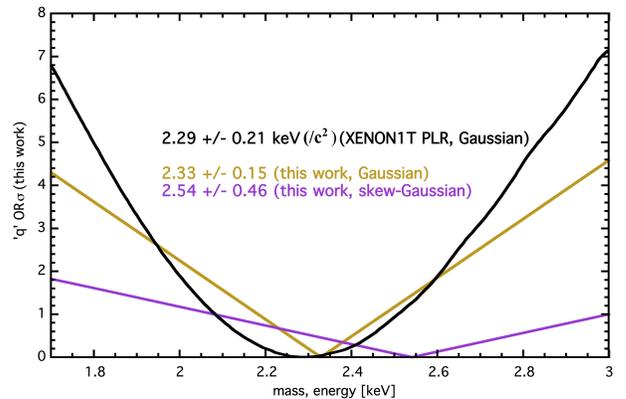}
\vspace{-10pt}
\caption{XENON1T's quoted log-likelihood ratio for different bosonic WIMP rest-mass energies from~\cite{Aprile:2020tmw} in black. The best fit was 2.3~keV. In gold, for comparison is the number of $\sigma$ of disagreement from $\chi^{2}$-based, not PLR, fits to the data, with NEST. The nature of the different statistical test causes the nonsmooth V shape, and lower significances of discrepancy in the ``wings,'' as expected for this type of method. Despite these differences, we find a near-identical best-fit energy as XENON1T, with 2.82~keV discrepant by a similar amount: $>3\sigma$ at least. In purple, the fit function is changed to a skew Gaussian, lowering the disagreement to 0.6$\sigma$ and bringing the best fit closer (higher $E$). This too is natural, as the simulation showed positive skew, quantitatively confirmed with data, and a greater number of free parameters introduces new correlations.}
\vspace{-20pt}
\label{BestFitBowl}
\end{center}
\end{figure}

Where $A$ is amplitude, $\mu$ mean, $\omega$ related to $\sigma$ (i.e., a measure of the width) and $\alpha$ is related to the amount of skew. $\xi$ can be either lower or higher than the mean, peak, or median, for positive and negative skew, respectively. We scan over both normal Gaussian fits (in gold) and skew versions (in purple) in the data in our Fig.~\ref{excess} (XENON1T's Fig.~4), conducting a monoenergetic peak search. The results are depicted in Fig.~\ref{BestFitBowl}. For the Gaussian case, we reproduce XENON1T's 2.3~keV value, with a similar error bar, further evidence they fit to a Gaussian, but in the skew case we find a higher best-fit mean, 2.5~keV, within a greater error range, spanning 2.3 and 2.8~keV. Contrasting the two methods, one can see that proper accounting of skew can easily shift 2.3 to 2.8 keV. Our \argon hypothesis should therefore still be seriously considered. A PLR would likely have a more constraining uncertainty. Lastly, many phenomenological papers reinterpreting the excess~\cite{an2020new,bloch2020exploring,he2020eft,alonsolvarez2020hidden,anchordoqui2020anomalous} infer a 2.8~keV peak in independent analyses completely unrelated to NEST, or skew. This is additional evidence 2.8 is not unreasonable.

Recognizing \cite{Aprile:2020tmw} states \argon is unlikely, we sought additional validation beyond mean energy and S2-only. First, we refer back to Fig.~\ref{Eres} inset, which shows NEST's width (15.79\% at 2.8 keV) in cyan, better matching the digitized XENON1T \argon calibration data width (15.88\%) in yellow, compared with the XENON1T model (18.88\% at 2.8 keV) in black, implying a possible discrepancy in energy resolution.~\cite{bloch2020exploring}, an independent reanalysis of the XENON1T data, agrees with the lower resolution predicted by NEST. The XENON1T analysis~\cite{AxionTalk} states that the \argon calibration data show a resolution of 18.12\%; however, both NEST and our digitization of the real data agree upon 16\% rounded across 3 methods (Gauss, skew, raw $\sigma$). Our only explanation is a fit to only the right half of the peak yields 18.1\%, but this just underscores again our point that skew or asymmetry cannot be ignored.

\begin{figure}
\begin{center}
\includegraphics[width=0.4\textwidth,clip]{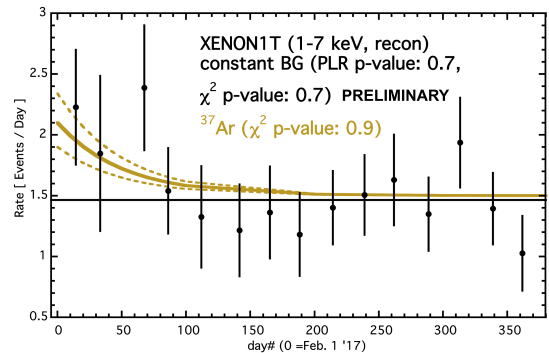}
\vspace{-10pt}
\caption{The time dependence of preliminary XENON1T science data~\cite{AxionTalk} in black. Without the time variation (flat black line), we find $p = 0.7$, matching the $p$ of the XENON1T PLR, in spite of using $\chi^{2}$-testing instead, showing the similarity of our analysis. We then introduce  31~$\pm$~11 \argon decays (yellow lines), as determined from fitting the excess in energy space, with no free parameters, then float the lifetime, counts, and both. The $\chi^{2}$/d.o.f for all scenarios is $<$~1.0; all hypotheses have corresponding p-values of $\sim$0.9. While this does not confirm $^{37}$Ar, it certainly does not rule it out either. Moreover, when the decay count is kept fixed at 31 \argon events, the best-fit mean ($1/e$) lifetime is 50 $^{+40}_{-30}$ days. When both the lifetime and counts are allowed to float free, the best fit is 57~$\pm$~31 events, and 36~$\pm~$21 days. $^{37}$Ar's actual lifetime is 50.6 days (half-life 35.04 d~\cite{Akimov_2014}). While errors on lifetime are large, multiple fit versions agreeing at $1\sigma$ is a positive hint and shows once more \argon is worth investigating thoroughly.}
\vspace{-25pt}
\label{TimeDep}
\end{center}
\end{figure}

Next, we considered time dependence in actual data in Fig.~\ref{TimeDep}. While errors are large and XENON1T's PLR has already established the points are consistent with none, we find in Fig.~\ref{TimeDep} statistical consistency with the \argon lifetime. While our hypothesis tests are ``goodness of fit'' not likelihood ratios, multiple tests all concur, despite distillation and gettering removing Ar in principle~\cite{Aprile:2020tmw}. The unlikely possibility exists that, e.g., a small leak, outgassing, or activation introduces minute quantities of it, or it is introduced by other means as-yet not understood. This could address why the excess was present in both of the two XENON1T science runs~\cite{Aprile:2020tmw}. This lifetime consistency implies an introduction mechanism occurring only at the beginning of runs. While we cannot explain conclusively why XENON1T would have 31 \argon events, we note LUX observed excess events at an energy consistent with $^{37}$Ar. If the LUX peak was new physics, XENON1T would have observed 200-500 events, based on the exposure increase between LUX and XENON1T, not 30~\cite{LUXAnnualMod}. This discrepancy cannot be accounted for by different efficiencies, since they were similar for both experiments ($\sim$100\% at 2.8~keV for ER).

\vspace{-15pt}
\section{Discussion}
\vspace{-5pt}
The excess seen by XENON1T can be effectively reproduced by NEST, and second it may be caused by known physics, other than tritium or other sources already considered~\cite{SenguptaChiSq}. On incorporation of \argon into the BG model, disagreement between model and data is $2.2\sigma$ (0.4$\sigma$ Poissonian). This is not completely comparable to PLR, but uses $\chi^{2}$, like~\cite{SenguptaChiSq}, but it might be possible to show even better agreement if we were to fully consider every uncertainty in NEST; we conservatively do not, relying on the default beta yields model.

There is uncertainty for the newly modeled skew~\cite{VetriLUX}. Advantage is never taken of this, using again the central NEST values only based on LUX/ZEPLIN~\cite{VetriLUX,LebedenkoPRD2009}. Higher skew, within error, could easily not only add more points at higher S2 in the first few S1 bins of Fig.~\ref{AllS2vS1Bands} but also add more counts into the 3.5 and 4.5~keV bins and make \argon as good if not a better fit to the XENON1T ER data, when compared again to the less well-motivated (from physics) exponential. That latter notion can itself still be motivated, based on past claims of new physics evidence~\cite{cogent} which may be explicable with exponential (or similar: power-law) rising backgrounds at low energy, across different technologies. We do not speculate on any specific physics to explain it in LXe.

Efficiency and energy reconstruction may contribute to systematics, but primarily at sub-keV; thus, these cannot impact the excess and overall XENON1T result. We acknowledge we had no access to actual XENON1T data and thus had to digitize their plots for comparisons. This can lead to a small error; although, NEST is incredibly robust in its predictions as depicted in the past, and we have put in a system of checks to try to minimize our errors. Therefore, the authors do not believe these would impact our reported results significantly. That said, and as mentioned before, NEST is an open-source software. We urge the XENON1T Collaboration to reproduce our work using their data and/or make their data available publicly. While the results presented here stop short of using PLR, such an analysis for the NEST results will yield more robust conclusions. Although, once more, it is unlikely to change the fact that to first order we have independently reproduced the XENON1T excess and find it consistent with $^{37}$Ar. We do not claim to know how it could be introduced, but note such an unexplained excess was previously found in LUX~\cite{LUXAnnualMod}.

Other possible future work could include redoing the entire analysis using the EXO-200 reported value $W=11.5$~eV, though this would be highly nontrivial: simple rescaling of $g_{1}$ and $g_{2}$ to account for this $W$ would disrupt NEST agreement with data on the carefully crafted fluctuations model (Fano factor for total quanta, excitation and ionization, and nonbinomial recombination fluctuations). Evidence in favor of our present assumptions ultimately lies in reproduction of XENON1T's data.

\vspace{-10pt}
\begin{acknowledgments}
\vspace{-10pt}
This work was supported by the University at Albany SUNY under new faculty startup funding for Prof.~Levy and by the DOE under Award~No.~DE-SC0015535. The authors wish to thank the LZ and LUX Collaborations for useful recent discussion as well as continued support for NEST work, plus their recognition of its high precision and NEST's extreme predictive power. Lastly, we wish to thank all NEST Collaboration members, especially those within XENON1T advocating for its increased usage.
\end{acknowledgments}
\vspace{-10pt}

\input{appendix.tex}


\input{output1112021.bbl}
\end{document}

%% file: appendix.tex
\appendix

\vspace{-0pt}
\section*{\label{sec:AppA}Appendix: Additional Validations}
\vspace{-5pt}
This Appendix is secondary evidence to corroborate several of our conclusions. First, we show that the photo-absorption process is capable of reducing the charge yield by $\sim$half (NEST actually assumes a smaller difference) compared to Compton scattering. The 2.8 keV \argon peak is the result of $e^-$ capture, so it was not immediately clear which of the two ER models was most appropriate, historically named gamma (photoabsorption would be better) and beta models, even though later data showed that betas agree with Compton scatters within uncertainties, in terms of yield measurements~\cite{Goetzke:2016lfg,Akerib:2019jtm}.

\begin{figure}[ht]
\begin{center}
\includegraphics[width=0.44\textwidth,clip]{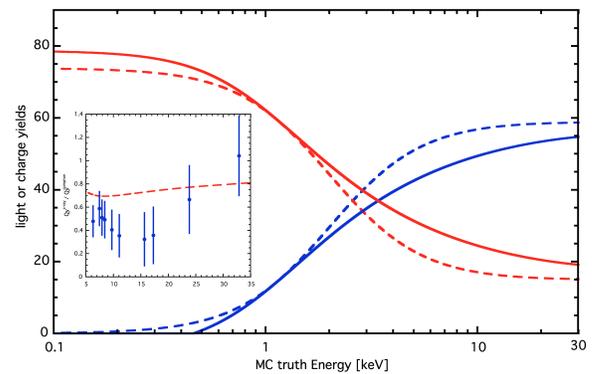}
\vspace{-5pt}
\caption{Comparison of NEST gamma (dash) and beta (solid) models below 30~keV, for 81~V/cm field. The light yield is in blue and charge yield in red. The inset depicts NEST's ratio of charge yields in red dash, along with a comparison to data, dividing the x-ray results of Ospanov and Obodovskii~\cite{obodovskii_ospanov} by the Compton scatters from Baudis~\textit{et~al.}, which also cites the former. Both data sets are from zero field, which is why NEST does not agree well with the data points despite being partly based on them. Direct evidence of the ratio at 81~V/cm does not exist. That being said, NEST is constrained by lower (0) and higher fields; its $\gamma$ model is extrapolated at nonzero field from high energy. This plot supplements Figs.~\ref{yields} and \ref{AllS2vS1Bands}, green.}
\vspace{-25pt}
\label{GammaProof}
\end{center}
\end{figure}

We base our claim of a difference primarily on \cite{Baudis_2013}. In the main text body, this is referred to as the difference between the nominal gamma/x-ray NEST model as opposed to the beta model which covers Compton as well. $L_{y}$ data were converted into $Q_{y}$ (even at 0~V/cm) by assuming anticorrelation holds (total of 73~quanta/keV). See Fig.~\ref{GammaProof}. Relative yields were converted into absolute numbers of photons per keV to high precision by converting between 32.1~keV ($^{83m}$Kr) and 122~keV ($^{57}$Co) yields, which are nearly identical~\cite{Kr83m}, and then assuming 63~$\pm$~2~photons/keV at 0~V/cm for $^{57}$Co $\gamma$-rays, a well-established value, given the historic role of this source in calibrating LXe detectors~\cite{NEST2015}. While many intermediate steps appear in this analysis, each is robustly justifiable.

If one reconsiders Fig.~\ref{yields}, different E-fields may be insufficient to completely explain at least $1\sigma$ of difference between the \argon PIXeY data \cite{pixey} and NEST. Recent work by XELDA~\cite{XELDA} indicates that there may be 5--10\% differences in yield at different energies and fields, not only between gammas and betas but among many different ER subtypes. The PIXeY data set most especially works in our favor here: if we increased the charge yield at 2.82~keV, it could better explain the excess observed in S2, at low S1s, in the data scatter plot of Fig.~\ref{AllS2vS1Bands} (around S1 of 7, S2 just below 2000~phe). This might further help explain reconstruction of 2.8~keV as 2.5 or as low as 2.3.

\vspace{-13pt}
\subsection{XENON1T's energy reconstruction}
\begin{figure}[ht]
\begin{center}
\includegraphics[width=0.4\textwidth,clip]{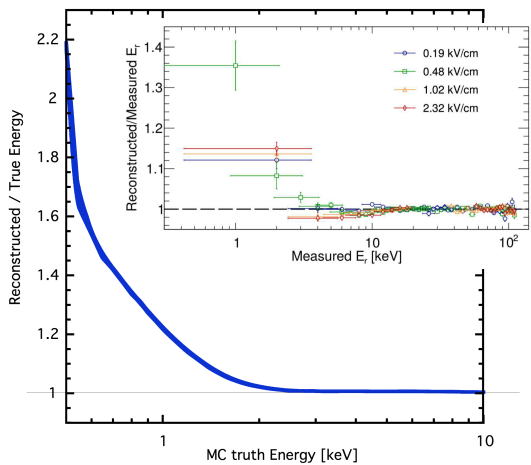}
\vspace{-10pt}
\caption{NEST output comparing true to reconstructed energy, using XENON1T parameters. The thickness of the line indicates statistical uncertainty. The disagreement is an emergent property stemming from many causes, including inherent skew in recombination probability, and triggering on upward fluctuations instead of true mean S1 and S2 pulse sizes, near thresholds (the Eddington bias~\cite{PhysRevLett.123.251801}). Inset: neriX data~\cite{Goetzke:2016lfg}, included for qualitative comparison only, as direct agreement would only be seen by modeling the neriX detector in NEST.}
\vspace{-20pt}
\label{ERecon}
\end{center}
\end{figure}

\vspace{-5pt}
As the excess was measured for the energy space histogram not in S2 versus S1 scatter, we also explored the energy reconstruction. While the combined-energy scale outperforms the older S1-only~\cite{XENONAxion2016} or ionization-only employed, e.g., by $\nu$ projects~\cite{microboone}, it is prone to breakdown at low energy. XENON1T reports \textit{reconstructed} energy, not true energy that they estimated via MC~\cite{PhysRevD.99.112009,Aprile2019:Rn220Fig16}. Figure~\ref{ERecon} shows the output from the NEST reconstructed energy, which differs drastically from the true energy especially in the sub-keV regime, in agreement with neriX~\cite{Goetzke:2016lfg}.

While important for other analyses, and although it can create differences of a factor of 2, the discrepancy is not relevant here. It is only particularly evident $<$1 keV, outside the region of interest for XENON1T's excess.

\vspace{-11pt}
\subsection{Detection efficiency}
\begin{figure}
\begin{center}
\includegraphics[width=0.5\textwidth,clip]{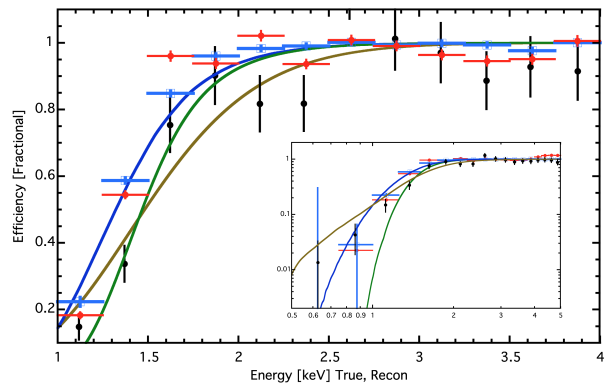}
\vspace{-15pt}
\caption{The dependence of the relative efficiency on the energy. The mustard is XENON1T's efficiency model and black is data, both from \cite{Aprile:2020tmw}, the latter using the \radon calibration, which we reproduce using NEST: first with a flat beta model (cyan) and then with the correct \radon energy spectrum (red). Red and cyan each follow black well: this provides further evidence we can replicate XENON1T's analyses. Green is NEST efficiency versus the reconstructed energy from an analytical fit (Gompertz) to a series of monoenergetic sims. Blue is versus true energy. Inset: zoom-out for larger range, log axes. An overall ($\sim$flat) reduction in efficiency across all energy is not portrayed, to focus on shape (actual asymptote $<$~1).}
\vspace{-20pt}
\label{DetEff}
\end{center}
\end{figure}

\vspace{-4pt}
All of the techniques for estimating detector efficiency ultimately agree on high efficiencies at 2--4~keV, of relevance to the excess. Despite not accounting for detector specifics such as unique S1 pulse shapes~\cite{Aprile2019:Rn220Fig16}, comparing NEST with data (Rn to Rn, red to black in Fig.~\ref{DetEff}), the reduced $\chi^{2}$ = 1.4 below 5~keV and 1.6 for 1--5~keV. These were calculated with systematics in both the data (Fig.~2 in~\cite{Aprile:2020tmw}) and in NEST (difference among red, cyan, green in Fig.~\ref{DetEff}). This points to NEST's robustness in modeling efficiency, even at energies of only a few keV.

Specifically, ER detection efficiency was verified in four ways: true energy for the x-axis (dark blue line), NEST reconstructed energy (green line) which should match the default XENON1T method (mustard line), simulating a flat energy spectrum (light blue, i.e., ``cyan'' points), and utilizing the \radon beta spectrum (red points), with the latter two but especially red meant to match black. The last three methods all use reconstructed energy, but differ in energy spectrum. Both mustard and black come from XENON1T: the former is their MC estimate and latter their Rn cross-check. NEST cases are compared to them.

Fig.~\ref{DetEff} demonstrates a good level of agreement among NEST's four scenarios, with the most significant comparisons being red and cyan against black, and green against mustard. Below 1~keV, mistaking the reconstructed energy for true (blue) may cause an overestimation of efficiency but this is challenging to conclude with great certainty given the large error bars including systematics. One of these systematics is the possibility that the ER light yield is higher than in NEST near 1~keV and lower energies, closer to what was assumed by XENON1T (Fig.~\ref{yields}) or in earlier NEST versions before sub-keV $Q_y$'s were published (driving $L_y$ estimates downward via anticorrelation). This could easily raise all NEST points and curves up to the mustard in the inset at the very lowest energies. While not directly relevant to the main point of this paper to explain the XENON1T excess since not in the $\sim$2--4~keV ROI or higher, we nonetheless continue to briefly discuss the region below 1 keV in this Appendix, as it may be of interest to the broader community.

The mustard line is above the black points for the first four bins in a row for the inset. However, the differences are always at $\sim$1--2$\sigma$. What we claim to be the efficiency vs.~true energy in (dark) blue is sometimes lower, sometimes higher, than the \radon points, but diverges from mustard as energy goes to zero. A continuous spectrum such as \radon is not best for determining efficiency, even though this was one LUX method~\cite{LUXtritium} (though not for a potential signal).~\radon was only the cross-check though for XENON1T. Alternatively, a dense series of monoenergetic MC peaks, as naturally done with NEST, can be tuned and verified to match a particular detector's data set, as performed for NRs for LUX~\cite{LUXPhD}. This should explain the difference between the green and mustard if the latter does not originate from a series of monoenergetic peaks. Much contamination between energy bins occurs due to finite resolution in continuous real data~\cite{jonthesis} that is of course changing rapidly versus energy, with resolution becoming poorer as energy decreases (Fig.~\ref{Eres}). If one prefers to study efficiency as a function of reconstructed energy with MC peaks instead of true, both mustard and black may be too high, above green. It is acceptable for black to disagree with green as \radon in black (in data) comes from a particular energy spectrum, but mustard and green should agree. Some difference in method and in $L_{y}$ have already been listed as the possible explanation. Individual PMT effects are another possibility: differing QEs and 2-PE probabilities by PMT across the detector, and XYZ-dependent photon collection. Additionally, the exact S1 pulse shape and area-dependent efficiency for detection of single-PE and/or few-PE pulses for the extreme energies, especially given XENON's uniquely long-tailed S1 pulses, with ringing, may be relevant.

Lastly, the green curve, the NEST efficiency versus reconstructed energy for XENON1T, does not agree better with the light blue (cyan) nor red points, also from NEST and also versus reconstructed energy, because it comes from the series of monoenergetic peaks, the results from which are effectively splined together, while red and cyan are influenced by cross-contamination between energy bins as mentioned earlier, from energies both higher and lower than the central value of a particular bin. The reason that red and cyan are not even self-consistent is the difference in energy spectrum (Rn versus flat). Rn is no longer approximately flat in energy near 1~keV.

%% file: output1112021.bbl
\providecommand{\noopsort}[1]{}\providecommand{\singleletter}[1]{#1}%

%% file: main.bbl
\begin{thebibliography}{62}%
\makeatletter
\providecommand \@ifxundefined [1]{%
 \@ifx{#1\undefined}
}%
\providecommand \@ifnum [1]{%
 \ifnum #1\expandafter \@firstoftwo
 \else \expandafter \@secondoftwo
 \fi
}%
\providecommand \@ifx [1]{%
 \ifx #1\expandafter \@firstoftwo
 \else \expandafter \@secondoftwo
 \fi
}%
\providecommand \natexlab [1]{#1}%
\providecommand \enquote  [1]{``#1''}%
\providecommand \bibnamefont  [1]{#1}%
\providecommand \bibfnamefont [1]{#1}%
\providecommand \citenamefont [1]{#1}%
\providecommand \href@noop [0]{\@secondoftwo}%
\providecommand \href [0]{\begingroup \@sanitize@url \@href}%
\providecommand \@href[1]{\@@startlink{#1}\@@href}%
\providecommand \@@href[1]{\endgroup#1\@@endlink}%
\providecommand \@sanitize@url [0]{\catcode `\\12\catcode `\$12\catcode
  `\&12\catcode `\#12\catcode `\^12\catcode `\_12\catcode `\%12\relax}%
\providecommand \@@startlink[1]{}%
\providecommand \@@endlink[0]{}%
\providecommand \url  [0]{\begingroup\@sanitize@url \@url }%
\providecommand \@url [1]{\endgroup\@href {#1}{\urlprefix }}%
\providecommand \urlprefix  [0]{URL }%
\providecommand \Eprint [0]{\href }%
\providecommand \doibase [0]{https://doi.org/}%
\providecommand \selectlanguage [0]{\@gobble}%
\providecommand \bibinfo  [0]{\@secondoftwo}%
\providecommand \bibfield  [0]{\@secondoftwo}%
\providecommand \translation [1]{[#1]}%
\providecommand \BibitemOpen [0]{}%
\providecommand \bibitemStop [0]{}%
\providecommand \bibitemNoStop [0]{.\EOS\space}%
\providecommand \EOS [0]{\spacefactor3000\relax}%
\providecommand \BibitemShut  [1]{\csname bibitem#1\endcsname}%
\let\auto@bib@innerbib\@empty
\bibitem [{\citenamefont {Rubin}(2000)}]{Rubin_2000}%
  \BibitemOpen
  \bibfield  {author} {\bibinfo {author} {\bibfnamefont {V.~C.}\ \bibnamefont
  {Rubin}},\ }\bibfield  {title} {\bibinfo {title} {One hundred years of
  rotating galaxies},\ }\href {https://doi.org/10.1086/316573} {\bibfield
  {journal} {\bibinfo  {journal} {Publications of the Astronomical Society of
  the Pacific}\ }\textbf {\bibinfo {volume} {112}},\ \bibinfo {pages} {747}
  (\bibinfo {year} {2000})}\BibitemShut {NoStop}%
\bibitem [{\citenamefont {Akrami}\ \emph {et~al.}(2020)\citenamefont {Akrami}
  \emph {et~al.}}]{planck2018}%
  \BibitemOpen
  \bibfield  {author} {\bibinfo {author} {\bibfnamefont {Y.}~\bibnamefont
  {Akrami}} \emph {et~al.} (\bibinfo {collaboration} {PLANCK Collaboration}),\
  }\bibfield  {title} {\bibinfo {title} {{Planck 2018 results. I. Overview and
  the cosmological legacy of Planck}},\ }\href
  {https://doi.org/10.1051/0004-6361/201833880} {\bibfield  {journal} {\bibinfo
   {journal} {Astronomy \& Astrophysics}\ }\textbf {\bibinfo {volume} {641}},\
  \bibinfo {pages} {A1} (\bibinfo {year} {2020})}\BibitemShut {NoStop}%
\bibitem [{\citenamefont {Peccei}\ and\ \citenamefont
  {Quinn}(1977)}]{PecceiQuinn}%
  \BibitemOpen
  \bibfield  {author} {\bibinfo {author} {\bibfnamefont {R.~D.}\ \bibnamefont
  {Peccei}}\ and\ \bibinfo {author} {\bibfnamefont {H.~R.}\ \bibnamefont
  {Quinn}},\ }\bibfield  {title} {\bibinfo {title} {$\mathrm{CP}$ conservation
  in the presence of pseudoparticles},\ }\href
  {https://doi.org/10.1103/PhysRevLett.38.1440} {\bibfield  {journal} {\bibinfo
   {journal} {Phys. Rev. Lett.}\ }\textbf {\bibinfo {volume} {38}},\ \bibinfo
  {pages} {1440} (\bibinfo {year} {1977})}\BibitemShut {NoStop}%
\bibitem [{\citenamefont {Aprile}\ \emph
  {et~al.}(2020{\natexlab{a}})\citenamefont {Aprile} \emph
  {et~al.}}]{Aprile:2020tmw}%
  \BibitemOpen
  \bibfield  {author} {\bibinfo {author} {\bibfnamefont {E.}~\bibnamefont
  {Aprile}} \emph {et~al.} (\bibinfo {collaboration} {XENON Collaboration}),\
  }\bibfield  {title} {\bibinfo {title} {{Excess electronic recoil events in
  XENON1T}},\ }\bibfield  {journal} {\bibinfo  {journal} {Physical Review D}\
  }\textbf {\bibinfo {volume} {102}},\ \href
  {https://doi.org/10.1103/physrevd.102.072004} {10.1103/physrevd.102.072004}
  (\bibinfo {year} {2020}{\natexlab{a}})\BibitemShut {NoStop}%
\bibitem [{\citenamefont {Szydagis}\ \emph {et~al.}(2020)\citenamefont
  {Szydagis} \emph {et~al.}}]{NESTsoftware}%
  \BibitemOpen
  \bibfield  {author} {\bibinfo {author} {\bibfnamefont {M.}~\bibnamefont
  {Szydagis}} \emph {et~al.},\ }\href@noop {} {\bibinfo {title} {{Open-access
  Noble Element Simulation Technique}}},\ \bibinfo {howpublished} {\url{
  https://zenodo.org/record/3905382#.XvlEAZNKjv1}} (\bibinfo {year}
  {2020})\BibitemShut {NoStop}%
\bibitem [{\citenamefont {Aprile}\ \emph
  {et~al.}(2019{\natexlab{a}})\citenamefont {Aprile} \emph
  {et~al.}}]{PhysRevD.99.112009}%
  \BibitemOpen
  \bibfield  {author} {\bibinfo {author} {\bibfnamefont {E.}~\bibnamefont
  {Aprile}} \emph {et~al.} (\bibinfo {collaboration} {XENON}),\ }\bibfield
  {title} {\bibinfo {title} {{XENON1T dark matter data analysis: Signal and
  background models and statistical inference}},\ }\href
  {https://doi.org/10.1103/PhysRevD.99.112009} {\bibfield  {journal} {\bibinfo
  {journal} {Phys. Rev. D}\ }\textbf {\bibinfo {volume} {99}},\ \bibinfo
  {pages} {112009} (\bibinfo {year} {2019}{\natexlab{a}})}\BibitemShut
  {NoStop}%
\bibitem [{\citenamefont {{Szydagis}}\ \emph {et~al.}(2011)\citenamefont
  {{Szydagis}}, \citenamefont {{Barry}}, \citenamefont {{Kazkaz}},
  \citenamefont {{Mock}}, \citenamefont {{Stolp}}, \citenamefont {{Sweany}},
  \citenamefont {{Tripathi}}, \citenamefont {{Uvarov}}, \citenamefont
  {{Walsh}},\ and\ \citenamefont {{Woods}}}]{NEST2011}%
  \BibitemOpen
  \bibfield  {author} {\bibinfo {author} {\bibfnamefont {M.}~\bibnamefont
  {{Szydagis}}}, \bibinfo {author} {\bibfnamefont {N.}~\bibnamefont {{Barry}}},
  \bibinfo {author} {\bibfnamefont {K.}~\bibnamefont {{Kazkaz}}}, \bibinfo
  {author} {\bibfnamefont {J.}~\bibnamefont {{Mock}}}, \bibinfo {author}
  {\bibfnamefont {D.}~\bibnamefont {{Stolp}}}, \bibinfo {author} {\bibfnamefont
  {M.}~\bibnamefont {{Sweany}}}, \bibinfo {author} {\bibfnamefont
  {M.}~\bibnamefont {{Tripathi}}}, \bibinfo {author} {\bibfnamefont
  {S.}~\bibnamefont {{Uvarov}}}, \bibinfo {author} {\bibfnamefont
  {N.}~\bibnamefont {{Walsh}}},\ and\ \bibinfo {author} {\bibfnamefont
  {M.}~\bibnamefont {{Woods}}},\ }\bibfield  {title} {\bibinfo {title} {{NEST:
  A Comprehensive Model for Scintillation Yield in Liquid Xenon}},\ }\href@noop
  {} {\bibfield  {journal} {\bibinfo  {journal} {JINST}\ }\textbf {\bibinfo
  {volume} {6}}\bibfield  {number} {\bibinfo  {number} { ({10})},\ \bibinfo
  {pages} {P10002}},\ }\Eprint {https://arxiv.org/abs/1106.1613}
  {arXiv:1106.1613 [physics.ins-det]} \BibitemShut {NoStop}%
\bibitem [{\citenamefont {Akerib}\ \emph {et~al.}(2014)\citenamefont {Akerib}
  \emph {et~al.}}]{LUX1}%
  \BibitemOpen
  \bibfield  {author} {\bibinfo {author} {\bibfnamefont {D.~S.}\ \bibnamefont
  {Akerib}} \emph {et~al.} (\bibinfo {collaboration} {LUX Collaboration}),\
  }\bibfield  {title} {\bibinfo {title} {{First Results from the LUX Dark
  Matter Experiment at the Sanford Underground Research Facility}},\ }\href
  {https://doi.org/10.1103/PhysRevLett.112.091303} {\bibfield  {journal}
  {\bibinfo  {journal} {Phys. Rev. Lett.}\ }\textbf {\bibinfo {volume} {112}},\
  \bibinfo {pages} {091303} (\bibinfo {year} {2014})}\BibitemShut {NoStop}%
\bibitem [{\citenamefont {Akerib}\ \emph
  {et~al.}(2020{\natexlab{a}})\citenamefont {Akerib} \emph {et~al.}}]{LZ}%
  \BibitemOpen
  \bibfield  {author} {\bibinfo {author} {\bibfnamefont {D.~S.}\ \bibnamefont
  {Akerib}} \emph {et~al.} (\bibinfo {collaboration} {LUX-ZEPLIN
  Collaboration}),\ }\bibfield  {title} {\bibinfo {title} {{Projected WIMP
  sensitivity of the LUX-ZEPLIN dark matter experiment}},\ }\href
  {https://doi.org/10.1103/PhysRevD.101.052002} {\bibfield  {journal} {\bibinfo
   {journal} {Phys. Rev. D}\ }\textbf {\bibinfo {volume} {101}},\ \bibinfo
  {pages} {052002} (\bibinfo {year} {2020}{\natexlab{a}})}\BibitemShut
  {NoStop}%
\bibitem [{\citenamefont {Ren}\ \emph {et~al.}(2018)\citenamefont {Ren} \emph
  {et~al.}}]{PANDAX}%
  \BibitemOpen
  \bibfield  {author} {\bibinfo {author} {\bibfnamefont {X.}~\bibnamefont
  {Ren}} \emph {et~al.} (\bibinfo {collaboration} {PandaX-II Collaboration}),\
  }\bibfield  {title} {\bibinfo {title} {{Constraining Dark Matter Models with
  a Light Mediator at the PandaX-II Experiment}},\ }\href
  {https://doi.org/10.1103/PhysRevLett.121.021304} {\bibfield  {journal}
  {\bibinfo  {journal} {Phys. Rev. Lett.}\ }\textbf {\bibinfo {volume} {121}},\
  \bibinfo {pages} {021304} (\bibinfo {year} {2018})}\BibitemShut {NoStop}%
\bibitem [{\citenamefont {Aprile}\ \emph {et~al.}(2016)\citenamefont {Aprile}
  \emph {et~al.}}]{XENON}%
  \BibitemOpen
  \bibfield  {author} {\bibinfo {author} {\bibfnamefont {E.}~\bibnamefont
  {Aprile}} \emph {et~al.} (\bibinfo {collaboration} {XENON Collaboration}),\
  }\bibfield  {title} {\bibinfo {title} {{Low-mass dark matter search using
  ionization signals in XENON100}},\ }\href
  {https://doi.org/10.1103/PhysRevD.94.092001} {\bibfield  {journal} {\bibinfo
  {journal} {Phys. Rev. D}\ }\textbf {\bibinfo {volume} {94}},\ \bibinfo
  {pages} {092001} (\bibinfo {year} {2016})}\BibitemShut {NoStop}%
\bibitem [{\citenamefont {Lenardo}\ \emph {et~al.}(2015)\citenamefont
  {Lenardo}, \citenamefont {Kazkaz}, \citenamefont {Manalaysay}, \citenamefont
  {Mock}, \citenamefont {Szydagis},\ and\ \citenamefont {Tripathi}}]{NEST2015}%
  \BibitemOpen
  \bibfield  {author} {\bibinfo {author} {\bibfnamefont {B.}~\bibnamefont
  {Lenardo}}, \bibinfo {author} {\bibfnamefont {K.}~\bibnamefont {Kazkaz}},
  \bibinfo {author} {\bibfnamefont {A.}~\bibnamefont {Manalaysay}}, \bibinfo
  {author} {\bibfnamefont {J.}~\bibnamefont {Mock}}, \bibinfo {author}
  {\bibfnamefont {M.}~\bibnamefont {Szydagis}},\ and\ \bibinfo {author}
  {\bibfnamefont {M.}~\bibnamefont {Tripathi}},\ }\bibfield  {title} {\bibinfo
  {title} {{A Global Analysis of Light and Charge Yields in Liquid Xenon}},\
  }\href {https://doi.org/10.1109/TNS.2015.2481322} {\bibfield  {journal}
  {\bibinfo  {journal} {IEEE Trans. Nucl. Sci.}\ }\textbf {\bibinfo {volume}
  {62}},\ \bibinfo {pages} {3387} (\bibinfo {year} {2015})},\ \Eprint
  {https://arxiv.org/abs/1412.4417} {arXiv:1412.4417 [astro-ph.IM]}
  \BibitemShut {NoStop}%
\bibitem [{\citenamefont {Cutter}(2017)}]{CutterTalk}%
  \BibitemOpen
  \bibfield  {author} {\bibinfo {author} {\bibfnamefont {J.}~\bibnamefont
  {Cutter}},\ }\href
  {http://nest.physics.ucdavis.edu/application/files/4315/3306/7234/cutter_nest_norcal_2017.pdf}
  {\bibinfo {title} {{The Noble Element Simulation Technique v2}}} (\bibinfo
  {year} {NorCal HEP-EXchange December 2, 2017})\BibitemShut {NoStop}%
\bibitem [{\citenamefont {Akerib}\ \emph
  {et~al.}(2018{\natexlab{a}})\citenamefont {Akerib} \emph {et~al.}}]{LUXg1g2}%
  \BibitemOpen
  \bibfield  {author} {\bibinfo {author} {\bibfnamefont {D.~S.}\ \bibnamefont
  {Akerib}} \emph {et~al.} (\bibinfo {collaboration} {LUX Collaboration}),\
  }\bibfield  {title} {\bibinfo {title} {{Calibration, event reconstruction,
  data analysis, and limit calculation for the LUX dark matter experiment}},\
  }\href {https://doi.org/10.1103/PhysRevD.97.102008} {\bibfield  {journal}
  {\bibinfo  {journal} {Phys. Rev. D}\ }\textbf {\bibinfo {volume} {97}},\
  \bibinfo {pages} {102008} (\bibinfo {year} {2018}{\natexlab{a}})}\BibitemShut
  {NoStop}%
\bibitem [{\citenamefont {Aprile}\ \emph
  {et~al.}(2018{\natexlab{a}})\citenamefont {Aprile} \emph {et~al.}}]{XENON3H}%
  \BibitemOpen
  \bibfield  {author} {\bibinfo {author} {\bibfnamefont {E.}~\bibnamefont
  {Aprile}} \emph {et~al.} (\bibinfo {collaboration} {XENON Collaboration}),\
  }\bibfield  {title} {\bibinfo {title} {{Signal yields of keV electronic
  recoils and their discrimination from nuclear recoils in liquid xenon}},\
  }\href {https://doi.org/10.1103/PhysRevD.97.092007} {\bibfield  {journal}
  {\bibinfo  {journal} {Phys. Rev. D}\ }\textbf {\bibinfo {volume} {97}},\
  \bibinfo {pages} {092007} (\bibinfo {year} {2018}{\natexlab{a}})}\BibitemShut
  {NoStop}%
\bibitem [{\citenamefont {Akerib}\ \emph {et~al.}(2019)\citenamefont {Akerib}
  \emph {et~al.}}]{Akerib:2019jtm}%
  \BibitemOpen
  \bibfield  {author} {\bibinfo {author} {\bibfnamefont {D.~S.}\ \bibnamefont
  {Akerib}} \emph {et~al.} (\bibinfo {collaboration} {LUX Collaboration}),\
  }\bibfield  {title} {\bibinfo {title} {{Improved Measurements of the
  $\beta$-Decay Response of Liquid Xenon with the LUX Detector}},\ }\href
  {https://doi.org/10.1103/PhysRevD.100.022002} {\bibfield  {journal} {\bibinfo
   {journal} {Phys. Rev. D}\ }\textbf {\bibinfo {volume} {100}},\ \bibinfo
  {pages} {022002} (\bibinfo {year} {2019})},\ \Eprint
  {https://arxiv.org/abs/1903.12372} {arXiv:1903.12372 [physics.ins-det]}
  \BibitemShut {NoStop}%
\bibitem [{\citenamefont {Boulton}\ \emph {et~al.}(2017)\citenamefont {Boulton}
  \emph {et~al.}}]{pixey}%
  \BibitemOpen
  \bibfield  {author} {\bibinfo {author} {\bibfnamefont {E.~M.}\ \bibnamefont
  {Boulton}} \emph {et~al.},\ }\bibfield  {title} {\bibinfo {title}
  {{Calibration of a two-phase xenon time projection chamber with a $^{37}$Ar
  source}},\ }\href {https://doi.org/10.1088/1748-0221/12/08/P08004} {\bibfield
   {journal} {\bibinfo  {journal} {JINST}\ }\textbf {\bibinfo {volume}
  {12}}\bibfield  {number} {\bibinfo  {number} { (08)},\ \bibinfo {pages}
  {P08004}},\ }\Eprint {https://arxiv.org/abs/1705.08958} {arXiv:1705.08958
  [physics.ins-det]} \BibitemShut {NoStop}%
\bibitem [{\citenamefont {Anton}\ \emph {et~al.}(2020)\citenamefont {Anton}
  \emph {et~al.}}]{EXO200WValue}%
  \BibitemOpen
  \bibfield  {author} {\bibinfo {author} {\bibfnamefont {G.}~\bibnamefont
  {Anton}} \emph {et~al.} (\bibinfo {collaboration} {EXO-200 Collaboration}),\
  }\bibfield  {title} {\bibinfo {title} {{Measurement of the scintillation and
  ionization response of liquid xenon at MeV energies in the EXO-200
  experiment}},\ }\href {https://doi.org/10.1103/PhysRevC.101.065501}
  {\bibfield  {journal} {\bibinfo  {journal} {Phys. Rev. C}\ }\textbf {\bibinfo
  {volume} {101}},\ \bibinfo {pages} {065501} (\bibinfo {year} {2020})},\
  \Eprint {https://arxiv.org/abs/1908.04128} {arXiv:1908.04128
  [physics.ins-det]} \BibitemShut {NoStop}%
\bibitem [{\citenamefont {Shockley}(2020)}]{AxionTalk}%
  \BibitemOpen
  \bibfield  {author} {\bibinfo {author} {\bibfnamefont {E.}~\bibnamefont
  {Shockley}},\ }\bibfield  {title} {\bibinfo {title} {{Search for New Physics
  with Electronic Recoil Events in XENON1T}},\ }\href
  {https://agenda.infn.it/event/23228/} {\bibfield  {journal} {\bibinfo
  {journal} {LNGS Seminar - https://agenda.infn.it/event/23228/}\ } (\bibinfo
  {year} {2020})}\BibitemShut {NoStop}%
\bibitem [{\citenamefont {Behrens}(2014)}]{Behrens}%
  \BibitemOpen
  \bibfield  {author} {\bibinfo {author} {\bibfnamefont {A.}~\bibnamefont
  {Behrens}},\ }\emph {\bibinfo {title} {Light Detectors for the XENON100 and
  XENON1T Dark Matter Search Experiments}},\ \href
  {https://www.physik.uzh.ch/groups/groupbaudis/darkmatter/theses/xenon/thesis_behrens.pdf}
  {Ph.D. thesis},\ \bibinfo  {school} {Universitaet Zurich} (\bibinfo {year}
  {2014})\BibitemShut {NoStop}%
\bibitem [{\citenamefont {Aprile}\ \emph
  {et~al.}(2018{\natexlab{b}})\citenamefont {Aprile} \emph
  {et~al.}}]{Aprile:2018dbl}%
  \BibitemOpen
  \bibfield  {author} {\bibinfo {author} {\bibfnamefont {E.}~\bibnamefont
  {Aprile}} \emph {et~al.} (\bibinfo {collaboration} {XENON Collaboration}),\
  }\bibfield  {title} {\bibinfo {title} {{Dark Matter Search Results from a One
  Ton-Year Exposure of XENON1T}},\ }\href
  {https://doi.org/10.1103/PhysRevLett.121.111302} {\bibfield  {journal}
  {\bibinfo  {journal} {Phys. Rev. Lett.}\ }\textbf {\bibinfo {volume} {121}},\
  \bibinfo {pages} {111302} (\bibinfo {year} {2018}{\natexlab{b}})},\ \Eprint
  {https://arxiv.org/abs/1805.12562} {arXiv:1805.12562 [astro-ph.CO]}
  \BibitemShut {NoStop}%
\bibitem [{\citenamefont {Aprile}\ \emph
  {et~al.}(2017{\natexlab{a}})\citenamefont {Aprile} \emph
  {et~al.}}]{Aprile:2017aty}%
  \BibitemOpen
  \bibfield  {author} {\bibinfo {author} {\bibfnamefont {E.}~\bibnamefont
  {Aprile}} \emph {et~al.} (\bibinfo {collaboration} {XENON Collaboration}),\
  }\bibfield  {title} {\bibinfo {title} {{The XENON1T Dark Matter
  Experiment}},\ }\href {https://doi.org/10.1140/epjc/s10052-017-5326-3}
  {\bibfield  {journal} {\bibinfo  {journal} {Eur. Phys. J. C}\ }\textbf
  {\bibinfo {volume} {77}},\ \bibinfo {pages} {881} (\bibinfo {year}
  {2017}{\natexlab{a}})},\ \Eprint {https://arxiv.org/abs/1708.07051}
  {arXiv:1708.07051 [astro-ph.IM]} \BibitemShut {NoStop}%
\bibitem [{\citenamefont {L\'opez~Paredes}\ \emph {et~al.}(2018)\citenamefont
  {L\'opez~Paredes}, \citenamefont {Ara\'ujo}, \citenamefont {Froborg},
  \citenamefont {Marangou}, \citenamefont {Olcina}, \citenamefont {Sumner},
  \citenamefont {Taylor}, \citenamefont {Tom\'as},\ and\ \citenamefont
  {Vacheret}}]{LOPEZPAREDES201856}%
  \BibitemOpen
  \bibfield  {author} {\bibinfo {author} {\bibfnamefont {B.}~\bibnamefont
  {L\'opez~Paredes}}, \bibinfo {author} {\bibfnamefont {H.}~\bibnamefont
  {Ara\'ujo}}, \bibinfo {author} {\bibfnamefont {F.}~\bibnamefont {Froborg}},
  \bibinfo {author} {\bibfnamefont {N.}~\bibnamefont {Marangou}}, \bibinfo
  {author} {\bibfnamefont {I.}~\bibnamefont {Olcina}}, \bibinfo {author}
  {\bibfnamefont {T.}~\bibnamefont {Sumner}}, \bibinfo {author} {\bibfnamefont
  {R.}~\bibnamefont {Taylor}}, \bibinfo {author} {\bibfnamefont
  {A.}~\bibnamefont {Tom\'as}},\ and\ \bibinfo {author} {\bibfnamefont
  {A.}~\bibnamefont {Vacheret}},\ }\bibfield  {title} {\bibinfo {title}
  {{Response of photomultiplier tubes to xenon scintillation light}},\ }\href
  {https://doi.org/10.1016/j.astropartphys.2018.04.006} {\bibfield  {journal}
  {\bibinfo  {journal} {Astropart. Phys.}\ }\textbf {\bibinfo {volume} {102}},\
  \bibinfo {pages} {56} (\bibinfo {year} {2018})},\ \Eprint
  {https://arxiv.org/abs/1801.01597} {arXiv:1801.01597 [physics.ins-det]}
  \BibitemShut {NoStop}%
\bibitem [{\citenamefont {Aprile}\ \emph
  {et~al.}(2017{\natexlab{b}})\citenamefont {Aprile} \emph
  {et~al.}}]{Aprile:2017iyp}%
  \BibitemOpen
  \bibfield  {author} {\bibinfo {author} {\bibfnamefont {E.}~\bibnamefont
  {Aprile}} \emph {et~al.} (\bibinfo {collaboration} {XENON}),\ }\bibfield
  {title} {\bibinfo {title} {{First Dark Matter Search Results from the XENON1T
  Experiment}},\ }\href {https://doi.org/10.1103/PhysRevLett.119.181301}
  {\bibfield  {journal} {\bibinfo  {journal} {Phys. Rev. Lett.}\ }\textbf
  {\bibinfo {volume} {119}},\ \bibinfo {pages} {181301} (\bibinfo {year}
  {2017}{\natexlab{b}})},\ \Eprint {https://arxiv.org/abs/1705.06655}
  {arXiv:1705.06655 [astro-ph.CO]} \BibitemShut {NoStop}%
\bibitem [{\citenamefont {Faham}\ \emph {et~al.}(2015)\citenamefont {Faham},
  \citenamefont {Gehman}, \citenamefont {Currie}, \citenamefont {Dobi},
  \citenamefont {Sorensen},\ and\ \citenamefont {Gaitskell}}]{2PE}%
  \BibitemOpen
  \bibfield  {author} {\bibinfo {author} {\bibfnamefont {C.}~\bibnamefont
  {Faham}}, \bibinfo {author} {\bibfnamefont {V.}~\bibnamefont {Gehman}},
  \bibinfo {author} {\bibfnamefont {A.}~\bibnamefont {Currie}}, \bibinfo
  {author} {\bibfnamefont {A.}~\bibnamefont {Dobi}}, \bibinfo {author}
  {\bibfnamefont {P.}~\bibnamefont {Sorensen}},\ and\ \bibinfo {author}
  {\bibfnamefont {R.}~\bibnamefont {Gaitskell}},\ }\bibfield  {title} {\bibinfo
  {title} {Measurements of wavelength-dependent double photoelectron emission
  from single photons in {VUV}-sensitive photomultiplier tubes},\ }\href
  {https://doi.org/10.1088/1748-0221/10/09/p09010} {\bibfield  {journal}
  {\bibinfo  {journal} {Journal of Instrumentation}\ }\textbf {\bibinfo
  {volume} {10}}\bibinfo  {number} { (2015)},\ \bibinfo {pages}
  {P09010}}\BibitemShut {NoStop}%
\bibitem [{\citenamefont {Akerib}\ \emph
  {et~al.}(2016{\natexlab{a}})\citenamefont {Akerib} \emph {et~al.}}]{LUXPhD}%
  \BibitemOpen
\bibfield  {number} {  }\bibfield  {author} {\bibinfo {author} {\bibfnamefont
  {D.~S.}\ \bibnamefont {Akerib}} \emph {et~al.} (\bibinfo {collaboration} {LUX
  Collaboration}),\ }\bibfield  {title} {\bibinfo {title} {{Improved Limits on
  Scattering of Weakly Interacting Massive Particles from Reanalysis of 2013
  LUX Data}},\ }\href {https://doi.org/10.1103/PhysRevLett.116.161301}
  {\bibfield  {journal} {\bibinfo  {journal} {Phys. Rev. Lett.}\ }\textbf
  {\bibinfo {volume} {116}},\ \bibinfo {pages} {161301} (\bibinfo {year}
  {2016}{\natexlab{a}})}\BibitemShut {NoStop}%
\bibitem [{\citenamefont {Akerib}\ \emph
  {et~al.}(2020{\natexlab{b}})\citenamefont {Akerib} \emph {et~al.}}]{LUXGreg}%
  \BibitemOpen
  \bibfield  {author} {\bibinfo {author} {\bibfnamefont {D.~S.}\ \bibnamefont
  {Akerib}} \emph {et~al.} (\bibinfo {collaboration} {LUX Collaboration}),\
  }\bibfield  {title} {\bibinfo {title} {{Improved modeling of $\beta$
  electronic recoils in liquid xenon using LUX calibration data}},\ }\href
  {https://doi.org/10.1088/1748-0221/15/02/T02007} {\bibfield  {journal}
  {\bibinfo  {journal} {JINST}\ }\textbf {\bibinfo {volume} {15}}\bibfield
  {number} {\bibinfo  {number} { (02)},\ \bibinfo {pages} {T02007}},\ }\Eprint
  {https://arxiv.org/abs/1910.04211} {arXiv:1910.04211 [physics.ins-det]}
  \BibitemShut {NoStop}%
\bibitem [{\citenamefont {Edwards}\ \emph {et~al.}(2018)\citenamefont {Edwards}
  \emph {et~al.}}]{pixey2}%
  \BibitemOpen
  \bibfield  {author} {\bibinfo {author} {\bibfnamefont {B.}~\bibnamefont
  {Edwards}} \emph {et~al.},\ }\bibfield  {title} {\bibinfo {title}
  {{Extraction efficiency of drifting electrons in a two-phase xenon time
  projection chamber}},\ }\href
  {https://doi.org/10.1088/1748-0221/13/01/P01005} {\bibfield  {journal}
  {\bibinfo  {journal} {JINST}\ }\textbf {\bibinfo {volume} {13}}\bibfield
  {number} {\bibinfo  {number} { (01)},\ \bibinfo {pages} {P01005}},\ }\Eprint
  {https://arxiv.org/abs/1710.11032} {arXiv:1710.11032 [physics.ins-det]}
  \BibitemShut {NoStop}%
\bibitem [{\citenamefont {Xu}\ \emph {et~al.}(2019)\citenamefont {Xu},
  \citenamefont {Pereverzev}, \citenamefont {Lenardo}, \citenamefont
  {Kingston}, \citenamefont {Naim}, \citenamefont {Bernstein}, \citenamefont
  {Kazkaz},\ and\ \citenamefont {Tripathi}}]{llnl}%
  \BibitemOpen
  \bibfield  {author} {\bibinfo {author} {\bibfnamefont {J.}~\bibnamefont
  {Xu}}, \bibinfo {author} {\bibfnamefont {S.}~\bibnamefont {Pereverzev}},
  \bibinfo {author} {\bibfnamefont {B.}~\bibnamefont {Lenardo}}, \bibinfo
  {author} {\bibfnamefont {J.}~\bibnamefont {Kingston}}, \bibinfo {author}
  {\bibfnamefont {D.}~\bibnamefont {Naim}}, \bibinfo {author} {\bibfnamefont
  {A.}~\bibnamefont {Bernstein}}, \bibinfo {author} {\bibfnamefont
  {K.}~\bibnamefont {Kazkaz}},\ and\ \bibinfo {author} {\bibfnamefont
  {M.}~\bibnamefont {Tripathi}},\ }\bibfield  {title} {\bibinfo {title}
  {{Electron extraction efficiency study for dual-phase xenon dark matter
  experiments}},\ }\href {https://doi.org/10.1103/PhysRevD.99.103024}
  {\bibfield  {journal} {\bibinfo  {journal} {Phys. Rev. D}\ }\textbf {\bibinfo
  {volume} {99}},\ \bibinfo {pages} {103024} (\bibinfo {year} {2019})},\
  \Eprint {https://arxiv.org/abs/1904.02885} {arXiv:1904.02885
  [physics.ins-det]} \BibitemShut {NoStop}%
\bibitem [{\citenamefont {Dahl}(2009)}]{Dahl:2009nta}%
  \BibitemOpen
  \bibfield  {author} {\bibinfo {author} {\bibfnamefont {C.~E.}\ \bibnamefont
  {Dahl}},\ }\emph {\bibinfo {title} {The physics of background discrimination
  in liquid xenon, and first results from XENON10 in the hunt for WIMP dark
  matter}},\ \href@noop {} {Ph.D. thesis},\ \bibinfo  {school} {Princeton
  University} (\bibinfo {year} {2009})\BibitemShut {NoStop}%
\bibitem [{\citenamefont {Goetzke}\ \emph {et~al.}(2017)\citenamefont
  {Goetzke}, \citenamefont {Aprile}, \citenamefont {Anthony}, \citenamefont
  {Plante},\ and\ \citenamefont {Weber}}]{Goetzke:2016lfg}%
  \BibitemOpen
  \bibfield  {author} {\bibinfo {author} {\bibfnamefont {L.}~\bibnamefont
  {Goetzke}}, \bibinfo {author} {\bibfnamefont {E.}~\bibnamefont {Aprile}},
  \bibinfo {author} {\bibfnamefont {M.}~\bibnamefont {Anthony}}, \bibinfo
  {author} {\bibfnamefont {G.}~\bibnamefont {Plante}},\ and\ \bibinfo {author}
  {\bibfnamefont {M.}~\bibnamefont {Weber}},\ }\bibfield  {title} {\bibinfo
  {title} {{Measurement of light and charge yield of low-energy electronic
  recoils in liquid xenon}},\ }\href
  {https://doi.org/10.1103/PhysRevD.96.103007} {\bibfield  {journal} {\bibinfo
  {journal} {Phys. Rev. D}\ }\textbf {\bibinfo {volume} {96}},\ \bibinfo
  {pages} {103007} (\bibinfo {year} {2017})},\ \Eprint
  {https://arxiv.org/abs/1611.10322} {arXiv:1611.10322 [astro-ph.IM]}
  \BibitemShut {NoStop}%
\bibitem [{\citenamefont {Aprile}\ \emph
  {et~al.}(2020{\natexlab{b}})\citenamefont {Aprile} \emph
  {et~al.}}]{Aprile:2020yad-ERes}%
  \BibitemOpen
  \bibfield  {author} {\bibinfo {author} {\bibfnamefont {E.}~\bibnamefont
  {Aprile}} \emph {et~al.} (\bibinfo {collaboration} {XENON}),\ }\bibfield
  {title} {\bibinfo {title} {{Energy resolution and linearity of XENON1T in the
  MeV energy range}},\ }\bibfield  {journal} {\bibinfo  {journal} {Eur. Phys.
  J. C}\ }\textbf {\bibinfo {volume} {80}},\ \href
  {https://doi.org/10.1140/epjc/s10052-020-8284-0}
  {10.1140/epjc/s10052-020-8284-0} (\bibinfo {year}
  {2020}{\natexlab{b}})\BibitemShut {NoStop}%
\bibitem [{\citenamefont {Akerib}\ \emph
  {et~al.}(2017{\natexlab{a}})\citenamefont {Akerib} \emph
  {et~al.}}]{LUXAgain}%
  \BibitemOpen
  \bibfield  {author} {\bibinfo {author} {\bibfnamefont {D.~S.}\ \bibnamefont
  {Akerib}} \emph {et~al.} (\bibinfo {collaboration} {LUX Collaboration}),\
  }\bibfield  {title} {\bibinfo {title} {Signal yields, energy resolution, and
  recombination fluctuations in liquid xenon},\ }\href
  {https://doi.org/10.1103/PhysRevD.95.012008} {\bibfield  {journal} {\bibinfo
  {journal} {Phys. Rev. D}\ }\textbf {\bibinfo {volume} {95}},\ \bibinfo
  {pages} {012008} (\bibinfo {year} {2017}{\natexlab{a}})}\BibitemShut
  {NoStop}%
\bibitem [{\citenamefont {Aprile}\ \emph
  {et~al.}(2019{\natexlab{b}})\citenamefont {Aprile} \emph
  {et~al.}}]{Aprile:2019dec}%
  \BibitemOpen
  \bibfield  {author} {\bibinfo {author} {\bibfnamefont {E.}~\bibnamefont
  {Aprile}} \emph {et~al.} (\bibinfo {collaboration} {XENON1T Collaboration}),\
  }\bibfield  {title} {\bibinfo {title} {{Observation of two-neutrino double
  electron capture in $^{124}$Xe with XENON1T}},\ }\href
  {https://doi.org/10.1038/s41586-019-1124-4} {\bibfield  {journal} {\bibinfo
  {journal} {Nature}\ }\textbf {\bibinfo {volume} {568}},\ \bibinfo {pages}
  {532} (\bibinfo {year} {2019}{\natexlab{b}})},\ \Eprint
  {https://arxiv.org/abs/1904.11002} {arXiv:1904.11002 [nucl-ex]} \BibitemShut
  {NoStop}%
\bibitem [{\citenamefont {Aprile}\ \emph
  {et~al.}(2019{\natexlab{c}})\citenamefont {Aprile} \emph
  {et~al.}}]{Aprile2019:Rn220Fig16}%
  \BibitemOpen
  \bibfield  {author} {\bibinfo {author} {\bibfnamefont {E.}~\bibnamefont
  {Aprile}} \emph {et~al.} (\bibinfo {collaboration} {XENON1T}),\ }\bibfield
  {title} {\bibinfo {title} {{XENON1T Dark Matter Data Analysis: Signal
  Reconstruction, Calibration and Event Selection}},\ }\href
  {https://doi.org/10.1103/PhysRevD.100.052014} {\bibfield  {journal} {\bibinfo
   {journal} {Phys. Rev. D}\ }\textbf {\bibinfo {volume} {100}},\ \bibinfo
  {pages} {052014} (\bibinfo {year} {2019}{\natexlab{c}})}\BibitemShut
  {NoStop}%
\bibitem [{\citenamefont {Lang}\ \emph {et~al.}(2016)\citenamefont {Lang},
  \citenamefont {Brown}, \citenamefont {Brown}, \citenamefont {Cervantes},
  \citenamefont {Macmullin}, \citenamefont {Masson}, \citenamefont
  {Schreiner},\ and\ \citenamefont {Simgen}}]{Lang:2016zde}%
  \BibitemOpen
  \bibfield  {author} {\bibinfo {author} {\bibfnamefont {R.~F.}\ \bibnamefont
  {Lang}}, \bibinfo {author} {\bibfnamefont {A.}~\bibnamefont {Brown}},
  \bibinfo {author} {\bibfnamefont {E.}~\bibnamefont {Brown}}, \bibinfo
  {author} {\bibfnamefont {M.}~\bibnamefont {Cervantes}}, \bibinfo {author}
  {\bibfnamefont {S.}~\bibnamefont {Macmullin}}, \bibinfo {author}
  {\bibfnamefont {D.}~\bibnamefont {Masson}}, \bibinfo {author} {\bibfnamefont
  {J.}~\bibnamefont {Schreiner}},\ and\ \bibinfo {author} {\bibfnamefont
  {H.}~\bibnamefont {Simgen}},\ }\bibfield  {title} {\bibinfo {title} {{A
  $^{220}$Rn source for the calibration of low-background experiments}},\
  }\href {https://doi.org/10.1088/1748-0221/11/04/P04004} {\bibfield  {journal}
  {\bibinfo  {journal} {JINST}\ }\textbf {\bibinfo {volume} {11}}\bibfield
  {number} {\bibinfo  {number} { (04)},\ \bibinfo {pages} {P04004}},\ }\Eprint
  {https://arxiv.org/abs/1602.01138} {arXiv:1602.01138 [physics.ins-det]}
  \BibitemShut {NoStop}%
\bibitem [{\citenamefont {Szydagis}\ \emph {et~al.}(2013)\citenamefont
  {Szydagis}, \citenamefont {Fyhrie}, \citenamefont {Thorngren},\ and\
  \citenamefont {Tripathi}}]{NEST2013}%
  \BibitemOpen
  \bibfield  {author} {\bibinfo {author} {\bibfnamefont {M.}~\bibnamefont
  {Szydagis}}, \bibinfo {author} {\bibfnamefont {A.}~\bibnamefont {Fyhrie}},
  \bibinfo {author} {\bibfnamefont {D.}~\bibnamefont {Thorngren}},\ and\
  \bibinfo {author} {\bibfnamefont {M.}~\bibnamefont {Tripathi}},\ }\bibfield
  {title} {\bibinfo {title} {{Enhancement of NEST Capabilities for Simulating
  Low-Energy Recoils in Liquid Xenon}},\ }\href
  {https://doi.org/10.1088/1748-0221/8/10/C10003} {\bibfield  {journal}
  {\bibinfo  {journal} {JINST}\ }\textbf {\bibinfo {volume} {8}}\bibfield
  {number} {\bibinfo  {number} { (10)},\ \bibinfo {pages} {C10003}},\ }\Eprint
  {https://arxiv.org/abs/1307.6601} {arXiv:1307.6601 [physics.ins-det]}
  \BibitemShut {NoStop}%
\bibitem [{\citenamefont {Rischbieter}(ting)}]{GregGammaX}%
  \BibitemOpen
  \bibfield  {author} {\bibinfo {author} {\bibfnamefont {G.}~\bibnamefont
  {Rischbieter}},\ }\bibfield  {title} {\bibinfo {title} {{Background Modeling
  in the LUX Detector for an Effective Field Theory Dark Matter Search}},\
  }\href@noop {} {\bibfield  {journal} {\bibinfo  {journal}
  {http://meetings.aps.org/Meeting/APR20/Session/R13.2}\ } (\bibinfo {year}
  {April 2020 APS Meeting})}\BibitemShut {NoStop}%
\bibitem [{\citenamefont {Dobi}(2014)}]{attilathesis}%
  \BibitemOpen
  \bibfield  {author} {\bibinfo {author} {\bibfnamefont {A.}~\bibnamefont
  {Dobi}},\ }\emph {\bibinfo {title} {{Measurement of the Electron Recoil Band
  of the LUX Dark Matter Detector With a Tritium Calibration Source}}},\
  \href@noop {} {Ph.D. thesis},\ \bibinfo  {school} {University of Maryland
  College Park} (\bibinfo {year} {2014})\BibitemShut {NoStop}%
\bibitem [{\citenamefont {Akerib}\ \emph
  {et~al.}(2018{\natexlab{b}})\citenamefont {Akerib} \emph
  {et~al.}}]{LUXAnnualMod}%
  \BibitemOpen
  \bibfield  {author} {\bibinfo {author} {\bibfnamefont {D.~S.}\ \bibnamefont
  {Akerib}} \emph {et~al.} (\bibinfo {collaboration} {LUX}),\ }\bibfield
  {title} {\bibinfo {title} {{Search for annual and diurnal rate modulations in
  the LUX experiment}},\ }\href {https://doi.org/10.1103/PhysRevD.98.062005}
  {\bibfield  {journal} {\bibinfo  {journal} {Phys. Rev. D}\ }\textbf {\bibinfo
  {volume} {98}},\ \bibinfo {pages} {062005} (\bibinfo {year}
  {2018}{\natexlab{b}})},\ \Eprint {https://arxiv.org/abs/1807.07113}
  {arXiv:1807.07113 [astro-ph.CO]} \BibitemShut {NoStop}%
\bibitem [{\citenamefont {Akerib}\ \emph
  {et~al.}(2020{\natexlab{c}})\citenamefont {Akerib} \emph
  {et~al.}}]{VetriLUX}%
  \BibitemOpen
  \bibfield  {author} {\bibinfo {author} {\bibfnamefont {D.~S.}\ \bibnamefont
  {Akerib}} \emph {et~al.} (\bibinfo {collaboration} {LUX Collaboration}),\
  }\bibfield  {title} {\bibinfo {title} {Discrimination of electronic recoils
  from nuclear recoils in two-phase xenon time projection chambers},\
  }\bibfield  {journal} {\bibinfo  {journal} {Physical Review D}\ }\textbf
  {\bibinfo {volume} {102}},\ \href
  {https://doi.org/10.1103/physrevd.102.112002} {10.1103/physrevd.102.112002}
  (\bibinfo {year} {2020}{\natexlab{c}})\BibitemShut {NoStop}%
\bibitem [{\citenamefont {Boulton}(2019)}]{boulton}%
  \BibitemOpen
  \bibfield  {author} {\bibinfo {author} {\bibfnamefont {E.~M.}\ \bibnamefont
  {Boulton}},\ }\emph {\bibinfo {title} {{Applications of Two-Phase Xenon Time
  Projection Chambers: Searching for Dark Matter and Special Nuclear
  Materials}}},\ \href@noop {} {Ph.D. thesis},\ \bibinfo  {school} {Yale
  University} (\bibinfo {year} {2019})\BibitemShut {NoStop}%
\bibitem [{\citenamefont {Priel}\ \emph {et~al.}(2017)\citenamefont {Priel},
  \citenamefont {Rauch}, \citenamefont {Landsman}, \citenamefont {Manfredini},\
  and\ \citenamefont {Budnik}}]{ModelSafeGuard}%
  \BibitemOpen
  \bibfield  {author} {\bibinfo {author} {\bibfnamefont {N.}~\bibnamefont
  {Priel}}, \bibinfo {author} {\bibfnamefont {L.}~\bibnamefont {Rauch}},
  \bibinfo {author} {\bibfnamefont {H.}~\bibnamefont {Landsman}}, \bibinfo
  {author} {\bibfnamefont {A.}~\bibnamefont {Manfredini}},\ and\ \bibinfo
  {author} {\bibfnamefont {R.}~\bibnamefont {Budnik}},\ }\bibfield  {title}
  {\bibinfo {title} {{A model independent safeguard against background
  mismodeling for statistical inference}},\ }\href
  {https://doi.org/10.1088/1475-7516/2017/05/013} {\bibfield  {journal}
  {\bibinfo  {journal} {JCAP}\ }\textbf {\bibinfo {volume} {05}}\bibfield
  {number} {\bibinfo  {number} { (013)}},\ }\Eprint
  {https://arxiv.org/abs/1610.02643} {arXiv:1610.02643 [physics.data-an]}
  \BibitemShut {NoStop}%
\bibitem [{\citenamefont {Akerib}\ \emph
  {et~al.}(2017{\natexlab{b}})\citenamefont {Akerib} \emph
  {et~al.}}]{LUXComplete}%
  \BibitemOpen
  \bibfield  {author} {\bibinfo {author} {\bibfnamefont {D.~S.}\ \bibnamefont
  {Akerib}} \emph {et~al.} (\bibinfo {collaboration} {LUX}),\ }\bibfield
  {title} {\bibinfo {title} {{Results from a search for dark matter in the
  complete LUX exposure}},\ }\href
  {https://doi.org/10.1103/PhysRevLett.118.021303} {\bibfield  {journal}
  {\bibinfo  {journal} {Phys. Rev. Lett.}\ }\textbf {\bibinfo {volume} {118}},\
  \bibinfo {pages} {021303} (\bibinfo {year} {2017}{\natexlab{b}})},\ \Eprint
  {https://arxiv.org/abs/1608.07648} {arXiv:1608.07648 [astro-ph.CO]}
  \BibitemShut {NoStop}%
\bibitem [{\citenamefont {An}\ \emph {et~al.}(2020)\citenamefont {An},
  \citenamefont {Pospelov}, \citenamefont {Pradler},\ and\ \citenamefont
  {Ritz}}]{an2020new}%
  \BibitemOpen
  \bibfield  {author} {\bibinfo {author} {\bibfnamefont {H.}~\bibnamefont
  {An}}, \bibinfo {author} {\bibfnamefont {M.}~\bibnamefont {Pospelov}},
  \bibinfo {author} {\bibfnamefont {J.}~\bibnamefont {Pradler}},\ and\ \bibinfo
  {author} {\bibfnamefont {A.}~\bibnamefont {Ritz}},\ }\href@noop {} {\bibinfo
  {title} {{New limits on dark photons from solar emission and keV scale dark
  matter}}} (\bibinfo {year} {2020}),\ \Eprint
  {https://arxiv.org/abs/2006.13929} {arXiv:2006.13929 [hep-ph]} \BibitemShut
  {NoStop}%
\bibitem [{\citenamefont {Bloch}\ \emph {et~al.}(2020)\citenamefont {Bloch},
  \citenamefont {Caputo}, \citenamefont {Essig}, \citenamefont {Redigolo},
  \citenamefont {Sholapurkar},\ and\ \citenamefont
  {Volansky}}]{bloch2020exploring}%
  \BibitemOpen
  \bibfield  {author} {\bibinfo {author} {\bibfnamefont {I.~M.}\ \bibnamefont
  {Bloch}}, \bibinfo {author} {\bibfnamefont {A.}~\bibnamefont {Caputo}},
  \bibinfo {author} {\bibfnamefont {R.}~\bibnamefont {Essig}}, \bibinfo
  {author} {\bibfnamefont {D.}~\bibnamefont {Redigolo}}, \bibinfo {author}
  {\bibfnamefont {M.}~\bibnamefont {Sholapurkar}},\ and\ \bibinfo {author}
  {\bibfnamefont {T.}~\bibnamefont {Volansky}},\ }\bibfield  {title} {\bibinfo
  {title} {{Exploring New Physics with O(keV) Electron Recoils in Direct
  Detection Experiments}},\ }\href@noop {} {\bibfield  {journal} {\bibinfo
  {journal} {pre-print}\ } (\bibinfo {year} {2020})},\ \Eprint
  {https://arxiv.org/abs/2006.14521} {arXiv:2006.14521 [hep-ph]} \BibitemShut
  {NoStop}%
\bibitem [{\citenamefont {He}\ \emph {et~al.}(2020)\citenamefont {He},
  \citenamefont {Wang},\ and\ \citenamefont {Zheng}}]{he2020eft}%
  \BibitemOpen
  \bibfield  {author} {\bibinfo {author} {\bibfnamefont {H.-J.}\ \bibnamefont
  {He}}, \bibinfo {author} {\bibfnamefont {Y.-C.}\ \bibnamefont {Wang}},\ and\
  \bibinfo {author} {\bibfnamefont {J.}~\bibnamefont {Zheng}},\ }\bibfield
  {title} {\bibinfo {title} {{EFT Analysis of Inelastic Dark Matter for Xenon
  Electron Recoil Detection}},\ }\href@noop {} {\bibfield  {journal} {\bibinfo
  {journal} {pre-print}\ } (\bibinfo {year} {2020})},\ \Eprint
  {https://arxiv.org/abs/2007.04963} {arXiv:2007.04963 [hep-ph]} \BibitemShut
  {NoStop}%
\bibitem [{\citenamefont {Alonso-\'Alvarez}\ \emph {et~al.}(2020)\citenamefont
  {Alonso-\'Alvarez}, \citenamefont {Ertas}, \citenamefont {Jaeckel},
  \citenamefont {Kahlhoefer},\ and\ \citenamefont
  {Thormaehlen}}]{alonsolvarez2020hidden}%
  \BibitemOpen
  \bibfield  {author} {\bibinfo {author} {\bibfnamefont {G.}~\bibnamefont
  {Alonso-\'Alvarez}}, \bibinfo {author} {\bibfnamefont {F.}~\bibnamefont
  {Ertas}}, \bibinfo {author} {\bibfnamefont {J.}~\bibnamefont {Jaeckel}},
  \bibinfo {author} {\bibfnamefont {F.}~\bibnamefont {Kahlhoefer}},\ and\
  \bibinfo {author} {\bibfnamefont {L.~J.}\ \bibnamefont {Thormaehlen}},\
  }\bibfield  {title} {\bibinfo {title} {{Hidden Photon Dark Matter in the
  Light of XENON1T and Stellar Cooling}},\ }\href@noop {} {\bibfield  {journal}
  {\bibinfo  {journal} {pre-print}\ } (\bibinfo {year} {2020})},\ \Eprint
  {https://arxiv.org/abs/2006.11243} {arXiv:2006.11243 [hep-ph]} \BibitemShut
  {NoStop}%
\bibitem [{\citenamefont {Anchordoqui}\ \emph {et~al.}(2020)\citenamefont
  {Anchordoqui}, \citenamefont {Antoniadis}, \citenamefont {Benakli},\ and\
  \citenamefont {Lüst}}]{anchordoqui2020anomalous}%
  \BibitemOpen
  \bibfield  {author} {\bibinfo {author} {\bibfnamefont {L.~A.}\ \bibnamefont
  {Anchordoqui}}, \bibinfo {author} {\bibfnamefont {I.}~\bibnamefont
  {Antoniadis}}, \bibinfo {author} {\bibfnamefont {K.}~\bibnamefont
  {Benakli}},\ and\ \bibinfo {author} {\bibfnamefont {D.}~\bibnamefont
  {Lüst}},\ }\bibfield  {title} {\bibinfo {title} {{Anomalous $U(1)$ gauge
  bosons as light dark matter in string theory}},\ }\href
  {https://doi.org/10.1016/j.physletb.2020.135838} {\bibfield  {journal}
  {\bibinfo  {journal} {Physics Letters B}\ }\textbf {\bibinfo {volume}
  {810}},\ \bibinfo {pages} {135838} (\bibinfo {year} {2020})}\BibitemShut
  {NoStop}%
\bibitem [{\citenamefont {Bhattacherjee}\ and\ \citenamefont
  {Sengupta}(2020)}]{SenguptaChiSq}%
  \BibitemOpen
  \bibfield  {author} {\bibinfo {author} {\bibfnamefont {B.}~\bibnamefont
  {Bhattacherjee}}\ and\ \bibinfo {author} {\bibfnamefont {R.}~\bibnamefont
  {Sengupta}},\ }\href@noop {} {\bibinfo {title} {{XENON1T Excess: Some
  Possible Backgrounds}}} (\bibinfo {year} {2020}),\ \Eprint
  {https://arxiv.org/abs/2006.16172} {arXiv:2006.16172 [hep-ph]} \BibitemShut
  {NoStop}%
\bibitem [{\citenamefont {Lebedenko}\ \emph {et~al.}(2009)\citenamefont
  {Lebedenko} \emph {et~al.}}]{LebedenkoPRD2009}%
  \BibitemOpen
  \bibfield  {author} {\bibinfo {author} {\bibfnamefont {V.~N.}\ \bibnamefont
  {Lebedenko}} \emph {et~al.} (\bibinfo {collaboration} {ZEPLIN-III
  Collaboration}),\ }\bibfield  {title} {\bibinfo {title} {{Results from the
  first science run of the ZEPLIN-III dark matter search experiment}},\
  }\bibfield  {journal} {\bibinfo  {journal} {Physical Review D}\ }\textbf
  {\bibinfo {volume} {80}},\ \href {https://doi.org/10.1103/physrevd.80.052010}
  {10.1103/physrevd.80.052010} (\bibinfo {year} {2009})\BibitemShut {NoStop}%
\bibitem [{\citenamefont {Aalseth}\ \emph {et~al.}(2011)\citenamefont {Aalseth}
  \emph {et~al.}}]{cogent}%
  \BibitemOpen
  \bibfield  {author} {\bibinfo {author} {\bibfnamefont {C.}~\bibnamefont
  {Aalseth}} \emph {et~al.} (\bibinfo {collaboration} {CoGeNT Collaboration}),\
  }\bibfield  {title} {\bibinfo {title} {{Results from a Search for Light-Mass
  Dark Matter with a P-type Point Contact Germanium Detector}},\ }\href
  {https://doi.org/10.1103/PhysRevLett.106.131301} {\bibfield  {journal}
  {\bibinfo  {journal} {Phys. Rev. Lett.}\ }\textbf {\bibinfo {volume} {106}},\
  \bibinfo {pages} {131301} (\bibinfo {year} {2011})},\ \Eprint
  {https://arxiv.org/abs/1002.4703} {arXiv:1002.4703 [astro-ph.CO]}
  \BibitemShut {NoStop}%
\bibitem [{\citenamefont {Akimov}\ \emph {et~al.}(2014)\citenamefont {Akimov}
  \emph {et~al.}}]{Akimov_2014}%
  \BibitemOpen
  \bibfield  {author} {\bibinfo {author} {\bibfnamefont {D.}~\bibnamefont
  {Akimov}} \emph {et~al.},\ }\bibfield  {title} {\bibinfo {title}
  {Experimental study of ionization yield of liquid xenon for electron recoils
  in the energy range 2.8--80 ke{V}},\ }\href
  {https://doi.org/10.1088/1748-0221/9/11/p11014} {\bibfield  {journal}
  {\bibinfo  {journal} {Journal of Instrumentation}\ }\textbf {\bibinfo
  {volume} {9}}\bibinfo  {number} { (2014)},\ \bibinfo {pages}
  {P11014}}\BibitemShut {NoStop}%
\bibitem [{\citenamefont {Baudis}\ \emph {et~al.}(2013)\citenamefont {Baudis},
  \citenamefont {Dujmovic}, \citenamefont {Geis}, \citenamefont {James},
  \citenamefont {Kish}, \citenamefont {Manalaysay}, \citenamefont
  {Marrodan~Undagoitia},\ and\ \citenamefont {Schumann}}]{Baudis_2013}%
  \BibitemOpen
\bibfield  {number} {  }\bibfield  {author} {\bibinfo {author} {\bibfnamefont
  {L.}~\bibnamefont {Baudis}}, \bibinfo {author} {\bibfnamefont
  {H.}~\bibnamefont {Dujmovic}}, \bibinfo {author} {\bibfnamefont
  {C.}~\bibnamefont {Geis}}, \bibinfo {author} {\bibfnamefont {A.}~\bibnamefont
  {James}}, \bibinfo {author} {\bibfnamefont {A.}~\bibnamefont {Kish}},
  \bibinfo {author} {\bibfnamefont {A.}~\bibnamefont {Manalaysay}}, \bibinfo
  {author} {\bibfnamefont {T.}~\bibnamefont {Marrodan~Undagoitia}},\ and\
  \bibinfo {author} {\bibfnamefont {M.}~\bibnamefont {Schumann}},\ }\bibfield
  {title} {\bibinfo {title} {{Response of liquid xenon to Compton electrons
  down to 1.5 keV}},\ }\href {https://doi.org/10.1103/PhysRevD.87.115015}
  {\bibfield  {journal} {\bibinfo  {journal} {Phys. Rev. D}\ }\textbf {\bibinfo
  {volume} {87}},\ \bibinfo {pages} {115015} (\bibinfo {year} {2013})},\
  \Eprint {https://arxiv.org/abs/1303.6891} {arXiv:1303.6891 [astro-ph.IM]}
  \BibitemShut {NoStop}%
\bibitem [{\citenamefont {Manalaysay}\ \emph {et~al.}(2010)\citenamefont
  {Manalaysay}, \citenamefont {Marrodan~Undagoitia}, \citenamefont {Askin},
  \citenamefont {Baudis}, \citenamefont {Behrens}, \citenamefont {Ferella},
  \citenamefont {Kish}, \citenamefont {Lebeda}, \citenamefont {Santorelli},
  \citenamefont {Venos},\ and\ \citenamefont {Vollhardt}}]{Kr83m}%
  \BibitemOpen
  \bibfield  {author} {\bibinfo {author} {\bibfnamefont {A.}~\bibnamefont
  {Manalaysay}}, \bibinfo {author} {\bibfnamefont {T.}~\bibnamefont
  {Marrodan~Undagoitia}}, \bibinfo {author} {\bibfnamefont {A.}~\bibnamefont
  {Askin}}, \bibinfo {author} {\bibfnamefont {L.}~\bibnamefont {Baudis}},
  \bibinfo {author} {\bibfnamefont {A.}~\bibnamefont {Behrens}}, \bibinfo
  {author} {\bibfnamefont {A.}~\bibnamefont {Ferella}}, \bibinfo {author}
  {\bibfnamefont {A.}~\bibnamefont {Kish}}, \bibinfo {author} {\bibfnamefont
  {O.}~\bibnamefont {Lebeda}}, \bibinfo {author} {\bibfnamefont
  {R.}~\bibnamefont {Santorelli}}, \bibinfo {author} {\bibfnamefont
  {D.}~\bibnamefont {Venos}},\ and\ \bibinfo {author} {\bibfnamefont
  {A.}~\bibnamefont {Vollhardt}},\ }\bibfield  {title} {\bibinfo {title}
  {{Spatially uniform calibration of a liquid xenon detector at low energies
  using $^{83m}$Kr}},\ }\bibfield  {journal} {\bibinfo  {journal} {Rev. Sci.
  Instrum.}\ }\textbf {\bibinfo {volume} {81}},\ \href
  {https://doi.org/10.1063/1.3436636} {10.1063/1.3436636} (\bibinfo {year}
  {2010})\BibitemShut {NoStop}%
\bibitem [{\citenamefont {Obodovskii}\ and\ \citenamefont
  {Ospanov}(1994)}]{obodovskii_ospanov}%
  \BibitemOpen
  \bibfield  {author} {\bibinfo {author} {\bibfnamefont {I.}~\bibnamefont
  {Obodovskii}}\ and\ \bibinfo {author} {\bibfnamefont {K.}~\bibnamefont
  {Ospanov}},\ }\bibfield  {title} {\bibinfo {title} {{Scintillation output of
  liquid xenon for low-energy $\gamma$-quanta}},\ }\href@noop {} {\bibfield
  {journal} {\bibinfo  {journal} {Pribory I Tekhnika Eksperimenta (USSR)}\
  }\textbf {\bibinfo {volume} {26}},\ \bibinfo {pages} {42} (\bibinfo {year}
  {1994})}\BibitemShut {NoStop}%
\bibitem [{\citenamefont {Temples}(2019)}]{XELDA}%
  \BibitemOpen
  \bibfield  {author} {\bibinfo {author} {\bibfnamefont {D.}~\bibnamefont
  {Temples}},\ }\href
  {http://www-kam2.icrr.u-tokyo.ac.jp/indico/event/3/session/10/contribution/414/material/slides/0.pdf}
  {\bibinfo {title} {{Understanding neutrino background implications in LXe-TPC
  dark matter searches using $^{127}$Xe electron captures}}} (\bibinfo {year}
  {TAUP 2019})\BibitemShut {NoStop}%
\bibitem [{\citenamefont {Aprile}\ \emph {et~al.}(2014)\citenamefont {Aprile}
  \emph {et~al.}}]{XENONAxion2016}%
  \BibitemOpen
  \bibfield  {author} {\bibinfo {author} {\bibfnamefont {E.}~\bibnamefont
  {Aprile}} \emph {et~al.} (\bibinfo {collaboration} {XENON100
  Collaboration}),\ }\bibfield  {title} {\bibinfo {title} {{First axion results
  from the XENON100 experiment}},\ }\href
  {https://doi.org/10.1103/PhysRevD.90.062009} {\bibfield  {journal} {\bibinfo
  {journal} {Phys. Rev. D}\ }\textbf {\bibinfo {volume} {90}},\ \bibinfo
  {pages} {062009} (\bibinfo {year} {2014})}\BibitemShut {NoStop}%
\bibitem [{\citenamefont {Adams}\ \emph {et~al.}(2020)\citenamefont {Adams}
  \emph {et~al.}}]{microboone}%
  \BibitemOpen
  \bibfield  {author} {\bibinfo {author} {\bibfnamefont {C.}~\bibnamefont
  {Adams}} \emph {et~al.} (\bibinfo {collaboration} {MicroBooNE
  Collaboration}),\ }\bibfield  {title} {\bibinfo {title} {{Calibration of the
  charge and energy loss per unit length of the MicroBooNE liquid argon time
  projection chamber using muons and protons}},\ }\href
  {https://doi.org/10.1088/1748-0221/15/03/P03022} {\bibfield  {journal}
  {\bibinfo  {journal} {JINST}\ }\textbf {\bibinfo {volume} {15}}\bibfield
  {number} {\bibinfo  {number} { (03)},\ \bibinfo {pages} {P03022}},\ }\Eprint
  {https://arxiv.org/abs/1907.11736} {arXiv:1907.11736 [physics.ins-det]}
  \BibitemShut {NoStop}%
\bibitem [{\citenamefont {Aprile}\ \emph
  {et~al.}(2019{\natexlab{d}})\citenamefont {Aprile} \emph
  {et~al.}}]{PhysRevLett.123.251801}%
  \BibitemOpen
  \bibfield  {author} {\bibinfo {author} {\bibfnamefont {E.}~\bibnamefont
  {Aprile}} \emph {et~al.} (\bibinfo {collaboration} {XENON Collaboration}),\
  }\bibfield  {title} {\bibinfo {title} {{Light Dark Matter Search with
  Ionization Signals in XENON1T}},\ }\href
  {https://doi.org/10.1103/PhysRevLett.123.251801} {\bibfield  {journal}
  {\bibinfo  {journal} {Phys. Rev. Lett.}\ }\textbf {\bibinfo {volume} {123}},\
  \bibinfo {pages} {251801} (\bibinfo {year} {2019}{\natexlab{d}})}\BibitemShut
  {NoStop}%
\bibitem [{\citenamefont {Akerib}\ \emph
  {et~al.}(2016{\natexlab{b}})\citenamefont {Akerib} \emph
  {et~al.}}]{LUXtritium}%
  \BibitemOpen
  \bibfield  {author} {\bibinfo {author} {\bibfnamefont {D.~S.}\ \bibnamefont
  {Akerib}} \emph {et~al.} (\bibinfo {collaboration} {LUX Collaboration}),\
  }\bibfield  {title} {\bibinfo {title} {{Tritium calibration of the LUX dark
  matter experiment}},\ }\href {https://doi.org/10.1103/PhysRevD.93.072009}
  {\bibfield  {journal} {\bibinfo  {journal} {Phys. Rev. D}\ }\textbf {\bibinfo
  {volume} {93}},\ \bibinfo {pages} {072009} (\bibinfo {year}
  {2016}{\natexlab{b}})},\ \Eprint {https://arxiv.org/abs/1512.03133}
  {arXiv:1512.03133 [physics.ins-det]} \BibitemShut {NoStop}%
\bibitem [{\citenamefont {Balajthy}(2018)}]{jonthesis}%
  \BibitemOpen
  \bibfield  {author} {\bibinfo {author} {\bibfnamefont {J.}~\bibnamefont
  {Balajthy}},\ }\emph {\bibinfo {title} {{Purity Monitoring Techniques and
  Electronic Energy Deposition Properties in Liquid Xenon Time Projection
  Chambers}}},\ \href@noop {} {Ph.D. thesis},\ \bibinfo  {school} {University
  of Maryland College Park} (\bibinfo {year} {2018})\BibitemShut {NoStop}%
\end{thebibliography}
